\documentclass[preprint2]{aastex61}

\shorttitle{Machine Learning in Astronomy}

\begin{document}

\title{Machine Learning in Astronomy: a practical overview}

\author{Dalya Baron}
\email{dalyabaron@gmail.com}
\affil{School of Physics and Astronomy \\
Tel-Aviv University \\
Tel Aviv 69978, Israel}



\begin{abstract}

Astronomy is experiencing a rapid growth in data size and complexity. This change fosters the development of data-driven science as a useful companion to the common model-driven data analysis paradigm, where astronomers develop automatic tools to mine datasets and extract novel information from them. In recent years, machine learning algorithms have become increasingly popular among astronomers, and are now used for a wide variety of tasks. In light of these developments, and the promise and challenges associated with them, the IAC Winter School 2018 focused on big data in Astronomy, with a particular emphasis on machine learning and deep learning techniques. This document summarizes the topics of supervised and unsupervised learning algorithms presented during the school, and provides practical information on the application of such tools to astronomical datasets. In this document I cover basic topics in supervised machine learning, including selection and preprocessing of the input dataset, evaluation methods, and three popular supervised learning algorithms, Support Vector Machines, Random Forests, and shallow Artificial Neural Networks. My main focus is on unsupervised machine learning algorithms, that are used to perform cluster analysis, dimensionality reduction, visualization, and outlier detection. Unsupervised learning algorithms are of particular importance to scientific research, since they can be used to extract new knowledge from existing datasets, and can facilitate new discoveries.

\end{abstract}


\section{Context}\label{sec:intro}

Astronomical datasets are undergoing a rapid growth in size and complexity, thus introducing Astronomy to the era of big data science (e.g., \citealt{ball10, pesenson10}). This growth is a result of past, ongoing, and future surveys, that produce massive multi-temporal and multi-wavelength datasets, with a wealth of information to be extracted and analyzed. Such surveys include the Sloan Digital Sky Survey (SDSS; \citealt{york00}), which provided the community with multi-color images of $\sim 1/3$ of sky, and high-resolution spectra of millions of Galactic and extra-galactic objects. Pan-STARRS \citep{kaiser10} and the Zwicky Transient Facility \citep{bellm14} perform a systematic exploration of the variable sky, delivering time-series of numerous asteroids, variable stars, supernovae, active galactic nuclei, and more. Gaia \citep{gaia16} is charting the three-dimensional map of the Milky Way, and will provide accurate positional and radial velocity measurements for over a billion stars in our Galaxy and throughout the Local Group. Future surveys, e.g., DESI \citep{levi13}, SKA \citep{dewdney09}, and LSST \citep{ivezic08}, will increase the number of available objects and their measured properties by more than an order of magnitude. 

In light of this accelerated growth, astronomers are developing automated tools to detect, characterize, and classify objects using the rich and complex datasets gathered with the different facilities. Machine learning algorithms have gained increasing popularity among astronomers, and are widely used for a variety of tasks. 

Machine learning algorithms are generally divided into two groups. Supervised machine learning algorithms are used to learn a mapping from a set of features to a target variable, based on example input-output pairs provided by a human expert (see e.g., \citealt{connolly95, collister04, fiorentin07, mahabal08, daniel11, laurino11, luis11, bloom12, brescia12, richards12, krone_martins14, masci14, miller15, wright15, djorgovski16, dlsanto16, lochner16, castro18, naul18, dlsanto18, dlsanto18b, krone_martins18, zucker18, delli_veneri19, ishida19, mahabal19, norris19, reis19}). Unsupervised learning algorithms are used to learn complex relationships that exist in the dataset, without labels provided by an expert. These can roughly be divided into clustering, dimensionality reduction, and anomaly detection (e.g., \citealt{boroson92, protopapas06, dabrusco09, vanderplas09, almeida10, ascasibar11, dabrusco12, meusinger12, fustes13, krone_martins14b, baron15, hocking15, gianniotis16, nun16, polsterer16, baron17, reis18a, reis18b}). The latter algorithms are arguably more important for scientific research, since they can be used to extract new knowledge from existing datasets, and can potentially facilitate new discoveries.

In view of the shift in data analysis paradigms and associated challenges, the IAC Winter School 2018 focused on big data in Astronomy. It included both lectures and hands-on tutorials, which are publicly available through their website\footnote{\url{http://www.iac.es/winterschool/2018/}}. The school covered the following topics: (1) general overview on the use of machine learning techniques in Astronomy: past, present and perspectives, (2) data challenges and solutions in forthcoming surveys, (3) supervised learning: classification and regression, (4) unsupervised learning and dimensionality reduction techniques, and (5) shallow and deep neural networks. In this document I summarize the topics of supervised and unsupervised learning algorithms, with special emphasis on unsupervised techniques. This document is not intended to provide a rigorous statistical background, but rather to present practical information on popular machine learning algorithms and their application to astronomical datasets. Supervised learning algorithms are discussed in section \ref{sec:supervised}, with an emphasis on optimization (section \ref{sec:optimizing_algs}), input datasets (section \ref{sec:input_dataset}), and three popular algorithms: Support Vector Machine (section \ref{sec:svm}), Decision Trees and Random Forest (section \ref{sec:ensemble_methods}), and shallow Artificial Neural Networks (section \ref{sec:ann}). Unsupervised learning algorithms are discussed in section \ref{sec:unsupervised_algs}, in particular distance assignment (section \ref{sec:distances}), clustering algorithms (section \ref{sec:clustering}), dimensionality reduction algorithms (section \ref{sec:dim_red}), and anomaly detection algorithms (section \ref{sec:outliers}).

\section{Supervised Learning}\label{sec:supervised}

Supervised machine learning algorithms are used to learn a relationship between a set of measurements and a target variable using a set of provided examples. Once obtained, the relationship can be used to predict the target variable of previously-unseen data. The main difference between traditional model fitting techniques and supervised learning algorithms is that in traditional model fitting the model is predefined, while supervised learning algorithms construct the model according to the input dataset. Supervised learning algorithms can, by construction, describe very complex non-linear relations between the set of measurements and the target variable, and can therefore be superior to traditional algorithms that are based on fitting of predefined models.

In machine learning terminology, the dataset consists of objects, and each object has measured features and a target variable. In Astronomy, the objects are usually physical entities such as stars or galaxies, and their features are measured properties, such as spectra or light-curves, or various higher-level quantities derived from observations, such as a variability period or stellar mass. The type of target variable depends on the particular task. In a \emph{classification} task, the target variables are discrete (often called labels), for example, classification of spectra into stars or quasars. In a \emph{regression} task, the target variable is continuous, for example, redshift estimation using photometric measurements. 

Supervised learning algorithms often have \emph{model parameters} that are estimated from the data. These parameters are part of the model that is learned from the data, are often saved as part of the learned model, and are required by the model when making predictions. Examples of model parameters include: support vectors in Support Vector Machines, splitting features and thresholds in Decision Trees and Random Forest, and weights of Artificial Neural Networks. In addition, supervised learning algorithms often have \emph{model hyper-parameters}, which are external to the model and whose values and are often set using different heuristics. Examples of model hyper-parameters include: the kernel shape in Support Vector Machines, the number of trees in Random Forests, and the number of hidden layers in Artificial Neural Networks. 

The application of supervised learning algorithms is usually divided into three stages. In the \emph{training stage}, the model hyper-parameters are set, and the model and the model parameters are learned from a subset of the input dataset, called the \emph{training set}. In the \emph{validation stage}, the model hyper-parameters are optimized according to some predefined cost function, often using a different subset of the input dataset, called the \emph{validation set}. During the validation stage, the model training is carried out iteratively for many different choices of hyper-parameters, and the hyper-parameters that result in the best performance on the validation set are chosen. Finally, in the \emph{test stage}, the trained model is used to predict the target variable of a different subset of the input dataset, called the \emph{test set}. The latter stage is necessary in order to assess the performance of the trained model on a previously-unseen dataset, i.e., a subset of the input data that was not used during the training and validation stages, and can be used to compare different supervised learning algorithms. Once these stages are completed, the model can be used to predict the target variable of new, previously-unseen datasets. 

This section provides some basic principles of supervised machine learning, and presents several popular algorithms used in Astronomy. For a detailed review on the subject, see Biehl (2019).
The section starts by describing the cost functions that are usually used to optimize model hyper-parameters, assess the performance of the final model, and compare different supervised learning algorithms (section \ref{sec:optimizing_algs}). Then, section \ref{sec:input_dataset} gives additional details on the input dataset, in particular its partition to training, validation, and test sets, feature selection and normalization, and imbalanced datasets. Finally, three popular algorithms are presented: Support Vector Machine (section \ref{sec:svm}), Decision Trees and Random Forest (section \ref{sec:ensemble_methods}), and shallow Artificial Neural Networks (section \ref{sec:ann}).

\begin{figure}
\includegraphics[width=0.49\textwidth]{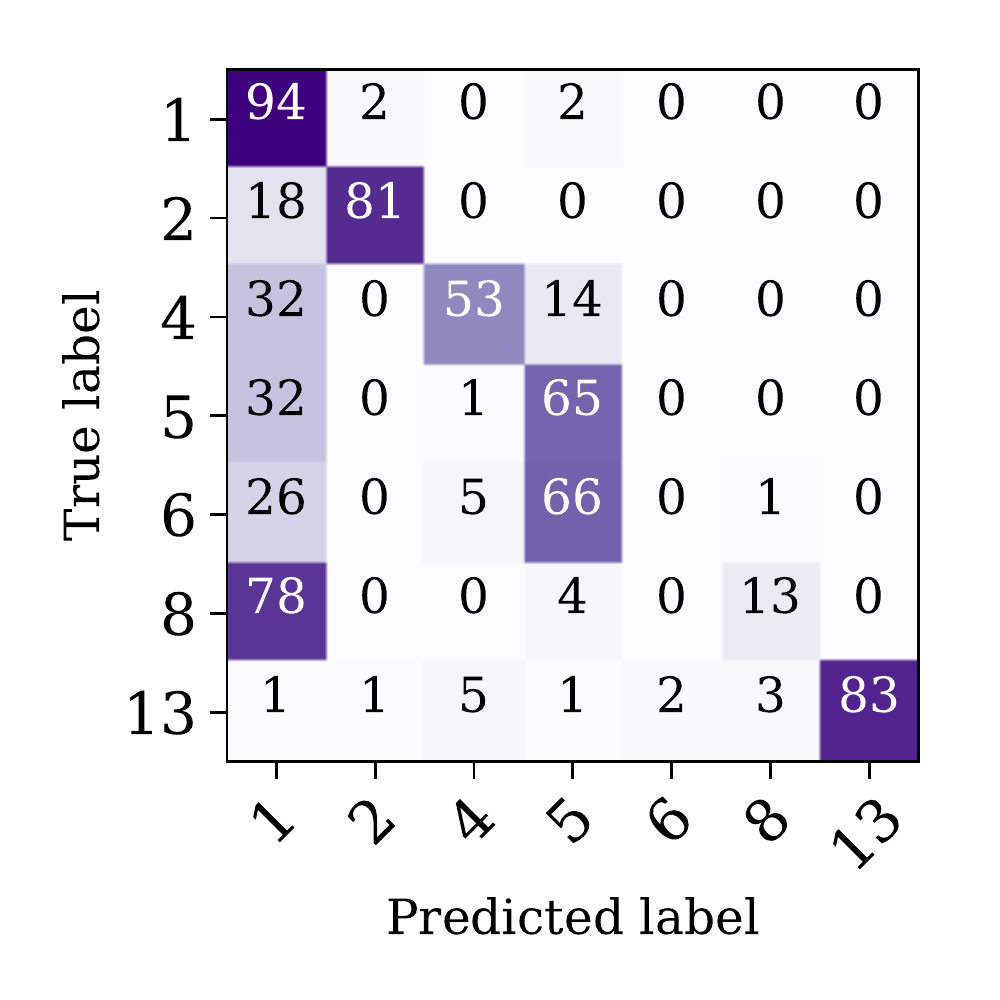}
\caption{Example of a confusion matrix taken from \citet{mahabal17}, who trained a deep learning model to distinguish between 7 classes of variable stars, marked by \texttt{1}, \texttt{2}, \texttt{4}, \texttt{5}, \texttt{6}, \texttt{8}, and \texttt{13} in the diagram. The confusion matrix shows the number of objects in each class versus the number of objects predicted by the model to belong to a particular class. In the best-case scenario, the confusion matrix will contain non-zero elements only in its diagonal, and zero elements otherwise.}
\label{f:confusion_matrix_example}
\end{figure}

\subsection{Evaluation Metrics}\label{sec:optimizing_algs}

There are different evaluation metrics one can use to optimize supervised learning algorithms. Evaluation metrics are used to optimize the model hyper-parameters, to assess the performance of the final model, to select optimal subset of features, and to compare between different supervised learning algorithms. The evaluation metrics are computed during the validation and the test stages, where the trained model is applied to a previously-unseen subset of the input dataset, and the target variable predicted by the model is compared to the target variable provided in the input data.

In regression tasks, where the target variable is continuous, the common metrics for evaluating the predictions of the model are the \emph{Mean Absolute Error} (MAE) and the \emph{Mean Squared Error} (MSE). The MAE is equal to $\frac{1}{n} \sum_{i=1}^{n}{| y_{i} - \hat{y}_{i} |}$, where $n$ is the number of objects in the validation or test set, $\hat{y}_{i}$ is the target variable predicted by the model, and $y_{i}$ is the target variable provided in the input data. The MSE is equal to $\frac{1}{n} \sum_{i=1}^{n}{( y_{i} - \hat{y}_{i} )^{2}}$. The MSE has the disadvantage of heavily weighting outliers, since it squares each term in the sum, giving outliers a larger weight. When outlier weighting is undesirable, it is better to use the MAE instead. Finally, it is worth noting additional metrics used in the literature, for example, the \emph{Normalized Median Absolute Deviation}, the \emph{Continuous Rank Probability Score}, and the \emph{Probability Integral Transform} (see e.g., \citealt{dlsanto18, dlsanto18b}). 

In classification tasks, where the target variable is discrete, the common evaluation metrics are the \emph{Classification Accuracy}, the \emph{Confusion Matrix}, and the \emph{Area Under ROC Curve}. Classification accuracy is the ratio of the number of correct predictions (i.e., the class predicted by the model is similar to the class provided in the input dataset) to the total number of predictions made. This value is obviously bound between 0 and 1. The accuracy should be used when the number of objects in each class is roughly similar, and when all predictions and prediction errors are equally important. Confusion matrices are used in classification tasks with more than two classes. Figure \ref{f:confusion_matrix_example} shows an example of a confusion matrix, taken from \citet{mahabal17}, who trained a supervised learning model to distinguish between 7 classes of variable stars. Each row and column in the matrix represents a particular class of objects, and the matrix shows the number of objects in each class versus the number of objects predicted by the model to belong to a particular class. In the best-case scenario, we expect the confusion matrix to be purely diagonal, with non-zero elements on the diagonal, and zero elements otherwise. Furthermore, similarly to the accuracy, one can normalize the confusion matrix to show values that range from 0 to 1, thus removing the dependence on the initial number of objects in each class. 

\begin{figure}
\includegraphics[width=0.49\textwidth]{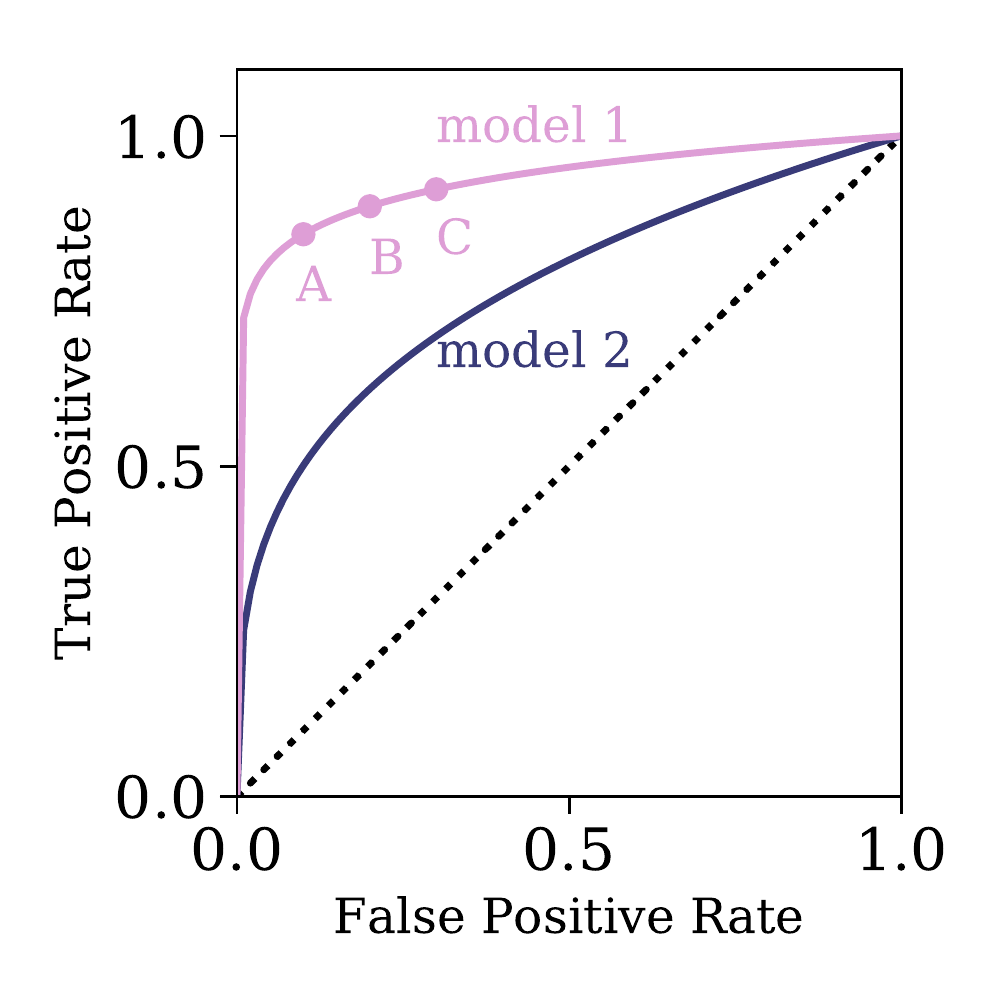}
\caption{An illustration of an ROC curve, where the true positive rate is plotted against the false positive rate. In the best-case scenario, we expect the true positive rate to be 1, and the false positive rate to be 0. The black line represents the resulting ROC curve for random assignment of classes, which is the worst-case scenario. The diagram is populated by varying the model hyper-parameters and plotting the true positive rate versus the false positive rate obtained for the validation set. The pink curve represents the ROC curve of model 1, where \texttt{A}, \texttt{B}, and \texttt{C} represent three particular choices of hyper-parameter value. The purple curve represents the ROC curve of model 2. The area under the ROC curve can be used to select the optimal model, which, in this case, is model 1.}
\label{f:roc_curve_example}
\end{figure}

Finally, the receiver operating characteristic curve (ROC curve) is a useful visualization tool of a supervised algorithm performance in a binary classification task. Figure \ref{f:roc_curve_example} shows an illustration of an ROC curve, where the \emph{True Positive Rate} is plotted against the \emph{False Positive Rate}. The true positive rate represents the number of "true" events that are correctly identified by the algorithm, divided by the total number of "true" events in the input dataset. The false positive rate represents the number of "false" events that were wrongly classified as "true" events divided by the total number of "false" events. For example, if we are interested in detecting gravitational lenses in galaxy images, "true" events are images with gravitational lenses, and "false" events are images without gravitational lenses. In the best-case scenario, we expect the true positive rate to be 1, and the false positive rate to be 0. The ROC curve diagram is generated by varying the model hyper-parameters, and plotting the true positive rate versus the false positive rate obtained for the validation set (the continuous ROC curves presented in Figure \ref{f:roc_curve_example} are produced by varying the classification/detection threshold of a particular algorithm. This threshold can be considered as a model hyper-parameter). Furthermore, the diagram can be used to compare the performance of different supervised learning algorithms, by selecting the algorithm with the maximal area under the curve. For example, in Figure \ref{f:roc_curve_example}, model 1 outperforms model 2 for any choice of hyper-parameters. Finally, depending on the task, one can decide to optimize differently. For some tasks one cannot tolerate false negatives (e.g., scanning for explosives in luggage), while for others the total number of errors is more important. 

\subsection{Input Dataset}\label{sec:input_dataset}

The input to any supervised learning algorithm consists of a set of objects with measured features, and a target variable which can be either continuous or discrete. As noted in the introduction to this section, the input dataset is divided into three sets, the training, validation, and test sets. The model is initially fit to the training set. Then, the model is applied to the validation set. The validation set provides an unbiased evaluation of the model performance while tuning the model's hyper-parameters. Validation sets are also used for regularization and to avoid overfitting, with the common practice of stopping the training process when the error on the validation dataset increases. Finally, the test set is used to provide an unbiased evaluation of the final model, and can be used to compare between different supervised learning algorithms. To perform a truly unbiased evaluation of the model performance, the training, validation, and test sets should be mutually exclusive. 

The dataset splitting ratios depend on the dataset size and on the algorithm one trains. Some algorithms require a substantial amount of data to train on, forcing one to enlarge the training set at the expense of the others. Algorithms with a few hyper-parameters, which are easily validated and tuned, require small validation sets, whereas models with many hyper-parameters might require larger validation sets. In addition, in \emph{Cross Validation} the dataset can be repeatedly split into training and validation sets, for example, by randomly selecting objects from a predefined set, and the model is then iteratively trained and validated on these different sets (see e.g., \citealt{miller17}). There are different splitting methods that are implemented in {\sc python} and are publicly available in the {\sc scikit-learn} library\footnote{\url{https://scikit-learn.org/stable/model_selection.html}}.

The performance of all supervised learning algorithms strongly depends on the input dataset, and in particular on the features that are selected to form the dataset. Most supervised learning algorithms are not constructed to work with hundreds or thousands of features, making feature selection a key part of the applied machine learning process. Feature selection can be done manually by an expert in the field, by defining features that are most probably relevant for the task at hand. There are various alternative ways to select an optimal set of features without domain knowledge, including filter methods, wrapper methods, and embedded methods. Filter methods assign a statistical score to each feature, and features are selected or removed according to this score. In wrapper methods, different combinations of features are prepared, and the combination that results in the best accuracy is selected. Embedded methods learn which features best contribute to the accuracy of the model during the model construction (see also \citealt{donalek13, dlsanto16, dlsanto18b}). Some to these methods are included in the {\sc scikit-learn} library\footnote{\url{https://scikit-learn.org/stable/modules/feature_selection.html}}. Finally, a pertinent note on deep learning, in particular \emph{Convolutional Neural Networks}. The structure of these networks allows them to take raw data as an input (e.g., spectra, light-curves, and images), and perform efficient feature extraction during the training process. Thus, using such models, there is usually no need to perform feature selection prior to the training stage.

Feature scaling is an additional key part of the data preparation process. While some algorithms do not require any feature scaling prior to training (e.g., Decision Trees and Random Forest), the performance of other algorithms strongly depends on it, and it is advised to apply some feature scaling prior to their training (e.g., for Support Vector Machine). There are various ways to scale the features in the initial dataset, including standardization, mean normalization, min-max scaling, and application of dimensionality reduction algorithms to the initial dataset. These will not be further discussed in this document, however, many feature scaling methods are available in {\sc scikit-learn}\footnote{\url{https://scikit-learn.org/stable/modules/preprocessing.html}}.

Finally, it is worth noting the topic of imbalanced datasets. Imbalanced data typically refers to the problem of a classification task where the classes are not equally represented. During training, many supervised learning algorithms assign equal weights to all the objects in the sample, resulting in a good performance on the larger class, and worse performance on the smaller class. In addition, the regular \emph{accuracy} cannot be used to evaluate the resulting model. Assume for example that we are interested in detecting gravitational lenses, and our dataset contains 100\,000 images of galaxies, out of which 100 images show evidence of lensing. For such a dataset, an algorithm that classifies \emph{all} objects as "not lens", regardless of the input features, will have an accuracy of 0.999. There are several methods to train and evaluate supervised learning algorithms in the presence of imbalanced datasets. One could apply different weights to different classes of objects during the training process, or undersample the larger class, or oversample the smaller class of objects. These, and additional methods, are available in {\sc scikit-learn}. Instead of the classification accuracy, one can use the ROC curve (figure \ref{f:roc_curve_example}), and select the hyper-parameters that result in the desired true positive versus false negative rates. 

\subsection{Support Vector Machine}\label{sec:svm}

\begin{figure}
\includegraphics[width=0.49\textwidth]{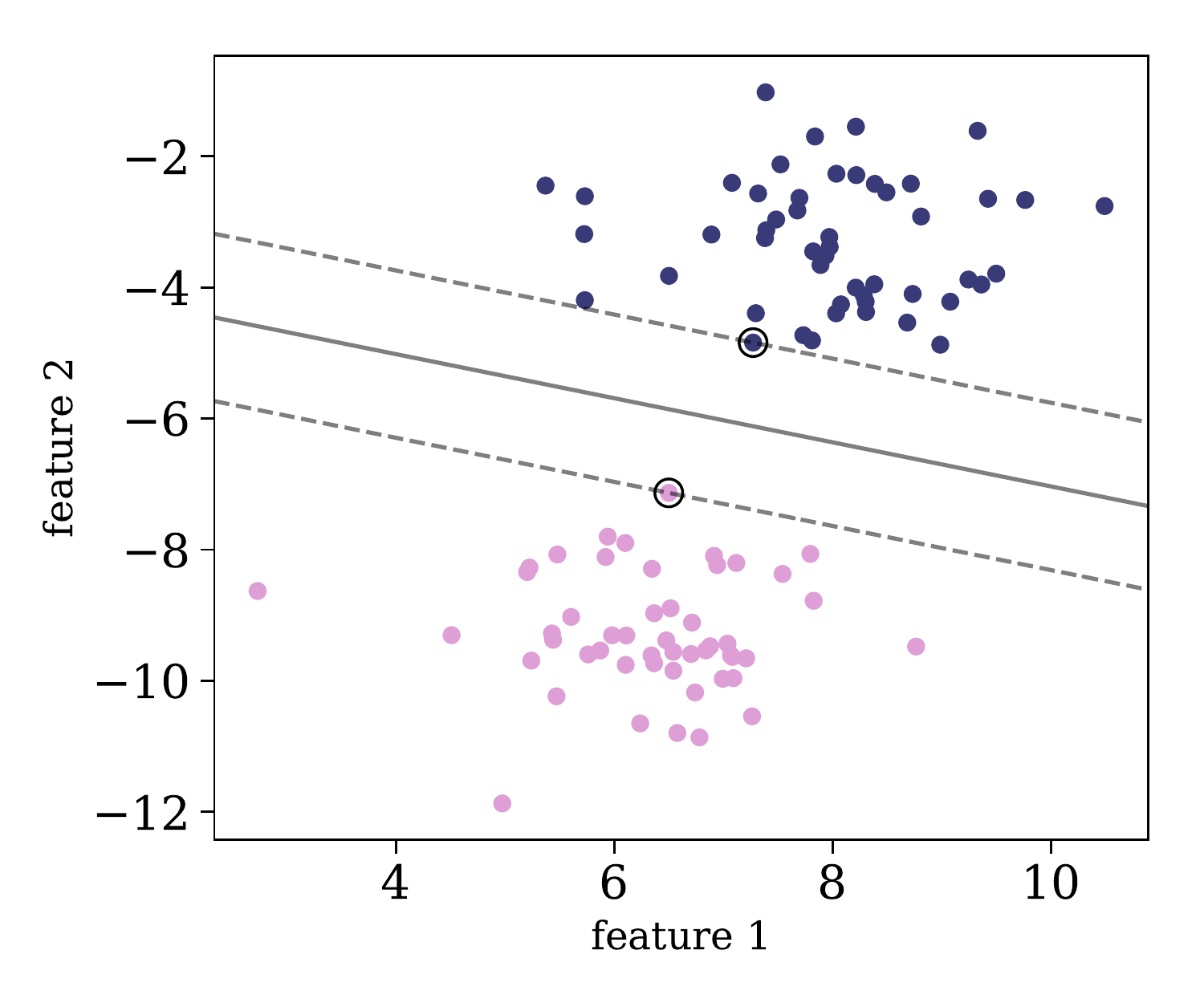}
\caption{Illustration of the SVM best hyperplane for a two-dimensional dataset with linearly-separable classes. The two classes are plotted with pink and purple circles, and the support vectors are marked with black circles. The hyperlane is marked with a solid grey line.}
\label{f:svm_linear_example}
\end{figure}

One of the most popular supervised learning algorithms is Support Vector Machine (SVM), which has been applied in Astronomy for a variety of tasks (e.g., \citealt{qu03, huertas_company08, fadely12, malek13, kovacs15, krakowski16, hartley17, hui18, ksoll18, pashchenko18}). Given a dataset with $N$ features, SVM finds a hyperplane in the $N$-dimensional space that best separates the given classes. In a two-dimensional space, this hyperplane is a line that divides the plane into two parts, where every class lies on a different side. The optimal hyperplane is defined to be the plane that has the maximal margin, i.e the maximum distance between the plane and the data points. The latter are called the support vectors. Once obtained, the hyperplane serves as a decision boundary, and new objects are classified according to their location with respect to the hyperplane. Figure \ref{f:svm_linear_example} shows an illustration of the SVM hyperplane for a two-dimensional dataset, in which the classes are linearly-separable. 

\begin{figure*}
	\centering
\includegraphics[width=0.98\textwidth]{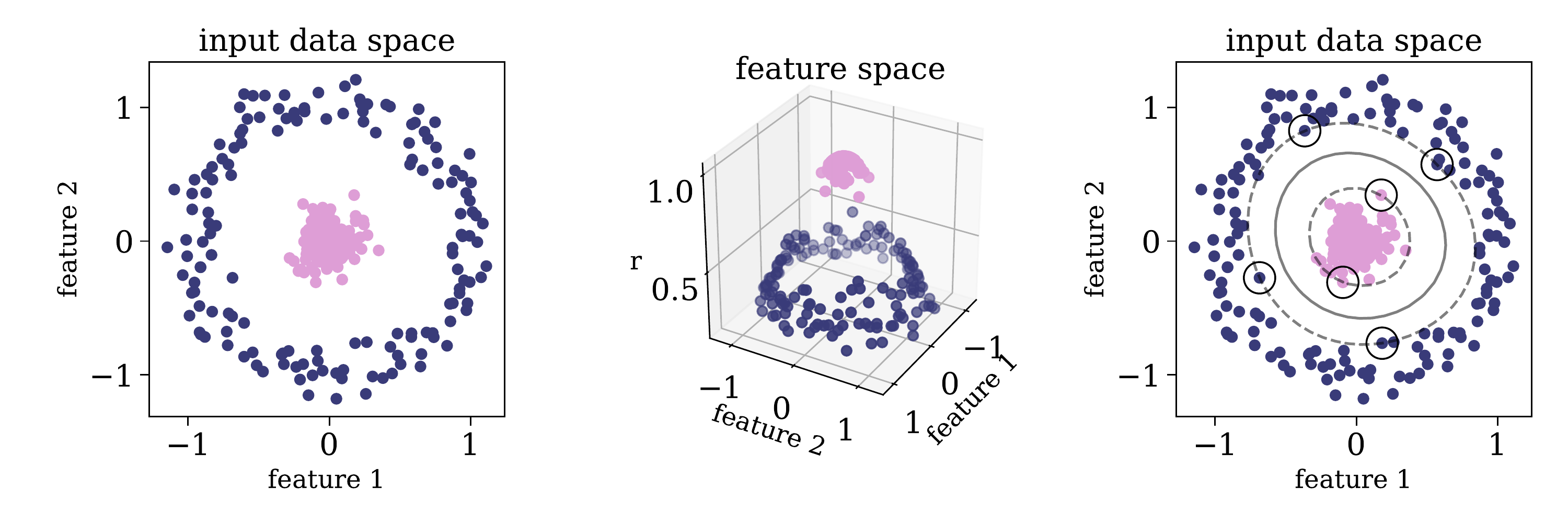}
\caption{Application of the SVM \emph{kernel trick} to a two-dimensional dataset that consists of two classes which are not linearly-separable. The left panel shows the dataset, where the different classes are represented by pink and purple circles. The middle panel shows the three-dimensional feature space that resulted from the applied mapping, where the classes can be separated with a two-dimensional hyperplane. The right panel shows the result of back-projecting the decision boundary to the input space, where the support vectors are marked with black circles, and the decision boundary with a solid grey line.}
\label{f:svm_non_linear_example}
\end{figure*}

More often than not, the different classes in the dataset are not linearly-separable. The left panel of figure \ref{f:svm_non_linear_example} shows an example of a two-dimensional dataset with two classes, which are not linearly-separable (i.e., the classes cannot be separated using a single straight line). The classification problem can be approached with the SVM \emph{kernel trick}. Instead of constructing the decision boundary in the input data space, the dataset is mapped into a transformed feature space, which is of higher dimension, where linear separation might be possible. Once the decision boundary is found, it is back-projected to the original input space, resulting in a non-linear boundary. The middle panel of figure \ref{f:svm_non_linear_example} shows the three-dimensional feature space that resulted from such a mapping, where, in this representation, the classes are linearly-separable. The right panel of figure \ref{f:svm_non_linear_example} shows the result of the back-projection of the decision boundary. To apply the kernel trick, one must define the \emph{kernel function} that is related to the non-linear feature mapping. There is a wide variety of kernel functions, the most popular being Gaussian Radial Basis Function (RBF), Polynomial, and Sigmoid. The kernel function is a hyper-parameter of SVM, and these functions usually depend on additional parameters, which are also hyper-parameters of the model. SVM is available in {\sc scikit-learn}\footnote{\url{https://scikit-learn.org/stable/modules/generated/sklearn.svm.SVC.html}}

SVM is simple and robust, allowing one to classify a large variety of datasets and to construct very non-linear decision boundaries. Since SVM is based on measuring Euclidean distances between the objects in the sample and the hyperplane, it is very sensitive to feature scaling. Therefore, it is advised to scale the features. SVM can be applied to datasets with many features, but its performance might be strongly affected by the presence of irrelevant features in the dataset. It is therefore recommended to perform feature selection prior to the training.

\subsection{Decision Trees and Random Forest}\label{sec:ensemble_methods}

Ensemble methods are meta-algorithms that combine several supervised learning techniques into a single predictive model, resulting in an overall improved performance, compared to the performance of each individual supervised algorithm. Ensemble methods either combine different supervised learning algorithms, or combine the information of a single algorithm that was trained on different subsets of the training set. One of the most popular ensemble methods is Random Forest, which is a collection of Decision Trees \citep{breiman84, breiman01}. Random Forest is mainly used as a supervised algorithm for classification and regression (e.g., \citealt{carliles10, bloom12, pichara12, pichara13, moller16, miller17, plewa18, yong18, ishida19}), but can also be used in an unsupervised setting, to produce similarity measures between the objects in the sample \citep{shi06, baron17, reis18a, reis18b}.	

\begin{figure}
\includegraphics[width=0.49\textwidth]{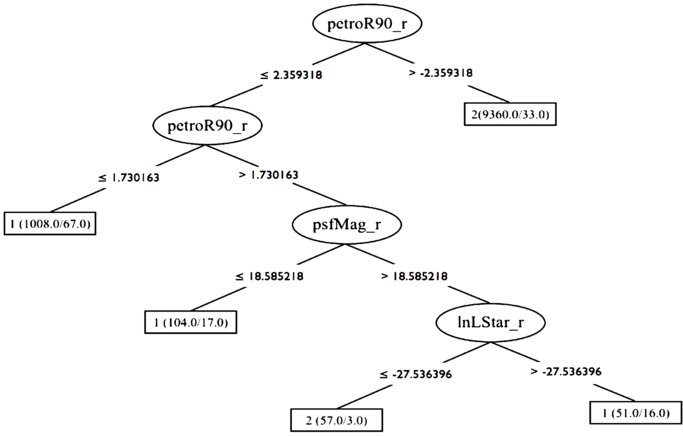}
\caption{An example of a trained decision tree, taken from \citet{vasconcellos11}. The decision tree contains nodes which represent conditions on features from the dataset, in this case  \texttt{petroR90\_r}, \texttt{psfMag\_r}, and \texttt{nLStar\_r}. The terminal nodes represent the assigned label of each particular path within the tree, which are \texttt{1}, \texttt{1}, \texttt{2}, \texttt{1}, and \texttt{2} from left to right, and represent stars and galaxies.}
\label{f:decision_tree_example}
\end{figure}

A decision tree is a non-parametric model constructed during the training stage, which is described by a top-to-bottom tree-like graph, and is used in both classification and regression tasks. The decision tree is a set of consecutive nodes, where each node represents a condition on one feature in the dataset. The conditions are of the form $X_{j} > X_{j, th}$, where $X_{j}$ is the value of the feature at index $j$ and $X_{j, th}$ is some threshold, both of which are determined during the training stage. The lowest nodes in the tree are called terminal nodes or leaves, and they do not represent a condition, but instead carry the assigned label of a particular path within the tree. Figure \ref{f:decision_tree_example} shows a visualization of a trained decision tree, taken from \citet{vasconcellos11}, who trained decision tree classifiers to distinguish stars from galaxies. 

To describe the training process of the decision tree, I consider the simple case of a classification task with two classes. The training process starts with the entire training set in the highest node of the tree -- the root. The algorithm searches for the feature $X_{j}$ and the feature threshold $X_{j,th}$ that result in the best separation of the two classes, where the definition of best separation is a model hyper-parameter, with the typical choices being the \emph{Gini impurity} or the \emph{information gain}. Once the best feature and best threshold are determined, the training set propagates to the left and right nodes below the root, according to the established condition. This process is repeated recursively, such that deeper nodes split generally smaller subsets of the original data. In its simplest version, the recursive process stops when every leaf of the tree contains a single class of objects. Once the decision tree is trained, it can be used to predict the class of previously-unseen objects. The prediction is done by propagating the object through the tree, according to its measured features and the conditions in the nodes. The predicted class of the object is then the label of the terminal leaf (for additional information see \citealt{breiman84, vasconcellos11, reis19}). 

Decision trees have several advantages. First, in their simpler forms, they have very few hyper-parameters, and can be applied to large datasets with numerous features. Second, their recursive structures are easily interpretable, in particular, they can be used to determine the feature importance. Feature importance represents the relative importance of different features to the classification task at hand. Since the trees are constructed recursively with the aim of splitting the dataset into the predefined classes, features that are selected earlier in the training process, closer to the root, are more important than features that are selected later, closer to the terminal nodes. Obviously, features that were not selected in any node during the training process, carry little relevant information to the classification task. In more complex versions, decision trees can provide a measure of uncertainty for the predicted classes (see e.g., \citealt{breiman84, vasconcellos11}). However, in the simplest version, there are no restrictions on the number of nodes or the depth of the resulting tree, making the algorithm extremely sensitive to outliers. The resulting classifier will typically show a perfect performance on the training set, but a poor performance on new previously-unseen datasets. A single decision tree is typically prone to overfitting the training data, and cannot generalize to new datasets. Therefore, it is rarely used in its single form. 

Random Forest is a collection of decision trees, where different decision trees are trained on different randomly-selected subsets of the original training set, and during the training of each individual tree, random subsets of the features are used to construct the conditions in individual nodes. This randomness reduces the correlation between the different trees, resulting in somewhat different tree structures with different conditions in their nodes. The Random Forest prediction is an aggregate of individual predictions of the trees in the forest, in the form of a majority vote. That is, a previously-unseen object is propagated through the different trees, and its assigned label is the label reached in the majority of the trees. While a single decision tree tends to overfit the training data, the combination of many decision trees in the form of a Random Forest generalizes well to previously unseen datasets, resulting in a better performance (for additional information see \citealt{breiman01}). As previously noted, Random Forest is one of the most popular machine learning algorithms in Astronomy. 

Random Forest can be applied to datasets with thousands of features, with a moderate increase in running time. The algorithm has a handful of hyper-parameters, such as the number of trees in the forest and the number of randomly-selected features to consider in each node in each tree. In terms of performance, model complexity, and number of hyper-parameters, Random Forest is between SVM and Deep Neural Networks. The main disadvantage of Random Forest, which applies to most supervised learning algorithms, is its inability to take into account feature and label uncertainties. This topic is of particular interest to astronomers, and I discuss it below in section \ref{sec:prf}.

Finally, it is worth noting that ensemble methods rely on the construction of a \emph{diverse} collection of classifiers and aggregation of their predictions (see e.g., \citealt{kuncheva03}). Ensemble methods in the form of "bagging" tend to decrease the classification variance, while methods in the form of "boosting" tend to decrease the classification bias. Random Forest is a "bagging" ensemble method, where the ensemble consists of a diverse collection of individual trees that are trained on different subsets of the data. In "boosting", the decision trees are built sequentially, such that each tree is presented with training samples that the previous tree failed to classify. One of the most popular "boosting" methods in Astronomy is Adaboost \citep{freund97}. The three algorithms described in this section are available in the {\sc scikit-learn} library\footnote{\url{https://scikit-learn.org/stable/modules/generated/sklearn.tree.DecisionTreeClassifier.html}\\ \url{https://scikit-learn.org/stable/modules/generated/sklearn.ensemble.RandomForestClassifier.html} \\ \url{https://scikit-learn.org/stable/modules/generated/sklearn.ensemble.AdaBoostClassifier.html}}.

\subsubsection{Probabilistic Random Forest}\label{sec:prf}

While shown to be very useful for various tasks in Astronomy, many Machine Learning algorithms were not designed for astronomical datasets, which are noisy and have gaps. In particular, measured features typically have a wide range of uncertainty values, and these uncertainties are often not taken into account when training the model. Indeed, the performance of Machine Learning algorithms depends strongly on the signal-to-noise ratio of the objects in the sample, and a model optimized on a dataset with particular noise characteristics will fail on a similar dataset with different noise properties. Furthermore, while in computer vision the labels provided to the algorithm are considered to be "ground truth" (e.g., classification of cats and dogs in images), in Astronomy the labels might suffer from some level of ambiguity. For example, in a classification task of "real" versus "bogus" in transient detection on difference images (see e.g., \citealt{bloom12}), the labels in the training set are obtained from a manual classification of scientists and citizen-scientists. While some might classify a given event as "real", others may classify it as "bogus". In addition, labels in the training set could be the output of a different algorithm, which provides a label with an associated uncertainty. Such uncertainties are also not treated by most Machine Learning algorithms. 

Recently, \citet{reis19} modified the traditional Random Forest algorithm to take into account uncertainties in the measurements (i.e., features) as well as in the assigned class. The \emph{Probabilistic Random Forest} algorithm treats the features and the labels as random variables rather than deterministic quantities, where each random variable is represented by a probability distribution function, whose mean is the provided measurement and its variance is the provided uncertainty. Their tests showed that the Probabilistic Random Forest outperforms the traditional Random Forest when applied to datasets with various noise properties, with an improvement of up to 10\% in classification accuracy with noisy features, and up to 30\% with noisy labels. In addition to the dramatic improvement in classification accuracy, the Probabilistic Random Forest naturally copes with missing values in the data, and outperforms Random Forest when applied to a dataset with different noise characteristics in the training and test sets. The Probabilistic Random Forest was implemented in {\sc python} and is publicly available on {\sc github}\footnote{\url{https://github.com/ireis/PRF}}.

\subsection{Artificial Neural Networks}\label{sec:ann}

Artificial neural networks are a set of algorithms with structures that are vaguely inspired by the biological neural networks that constitute the human brain. Their flexible structure and non-linearity allows one to perform a wide variety of tasks, including classification and regression, clustering, and dimensionality reduction, making them extremely popular in Astronomy (e.g., \citealt{storrie_lombardi92, weaver95, singh98, snider01, firth03, tagliaferri03, vanzella04, blake07, banerji10, eatough10, brescia13, brescia14, ellison16, teimoorinia16, mahabal17, bilicki18, huertas_company18, naul18, parks18, das19}). In this section I describe the main building blocks of \emph{shallow} artificial neural networks, and their use for classification and regression tasks. The section does not include details on \emph{Deep Learning}, in particular \emph{Convolutional Neural Networks}, \emph{Recurrent Neural Networks}, and \emph{Generative Adversarial Networks} (see lectures by M. Huertas-Company for details on deep learning). 

\begin{figure}
\includegraphics[width=0.49\textwidth]{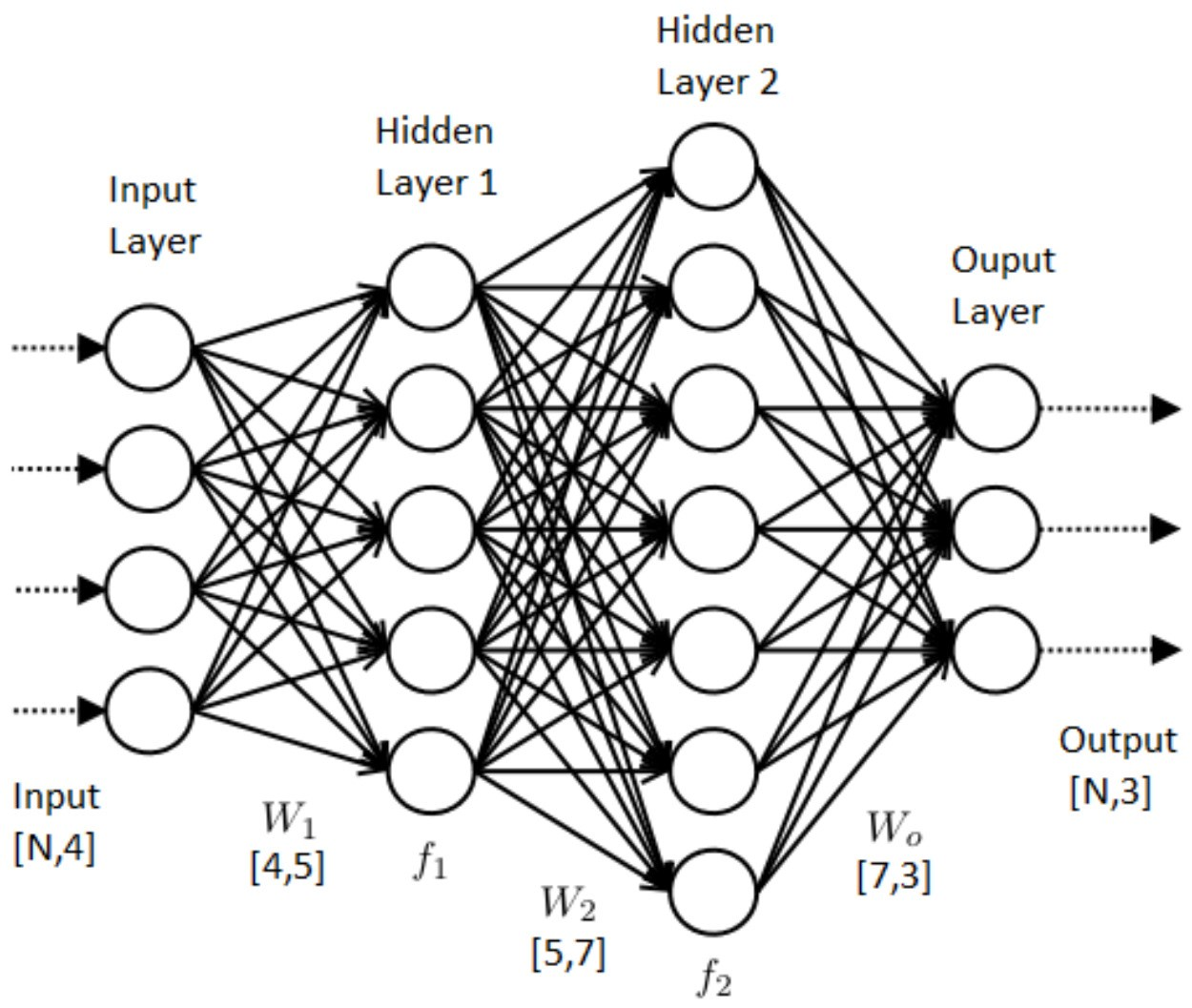}
\caption{Illustration of a shallow neural network architecture. The network consists of an input layer, two hidden layers, and an output layer. The input dataset is propagated from the input layer, through the hidden layers, to the output layer, where a prediction of a target variable is made. Each neuron is a linear combination of the neuron values in the previous layer, followed by an application of a non-linear activation function (see text for more details).}
\label{f:shallow_nn_example}
\end{figure}

Figure \ref{f:shallow_nn_example} is an illustration of a shallow neural network architecture. The network consists of an \emph{input layer}, \emph{output layer}, and several \emph{hidden layers}, where each of these contain neurons that transmit information to the neurons in the succeeding layer. The input data is transmitted from the input layer, through the hidden layers, and reaches the output layer, where the target variable is predicted. The value of every neuron in the network (apart from the neurons in the input layer) is a linear combination of the neurons in the previous layer, followed by an application of a non-linear activation function. That is, the values of the neurons in the first hidden layer are given by $\vec{x}_{1} = f_{1} \big( W_{1} \vec{x}_{0} \big)$, where $\vec{x}_{0}$ is a vector that describes the values of the neurons in the input layer (the input data), $W_{1}$ is a weight matrix that describes the linear combination of the input values, and $f_{1}$ is a non-linear activation function. In a similar manner, the values of the neurons in the second hidden layer are given by $\vec{x}_{2} = f_{2} \big( W_{2} \vec{x}_{1} \big)$, with a similar notation. Finally, the values of the neurons in the output layer are given by $\vec{x}_{3} = f_{3} \big( W_{3} \vec{x}_{2} \big) = f_{3} \big( W_{3} f_{2} \big( W_{2} f_{1} \big( W_{1} \vec{x}_{0} \big) \big) \big)$. The weights of the network are model parameters which are optimized during training via \emph{back-propagation} (for additional details see lectures by M. Huertas-Company). The non-linear activation functions are model hyper-parameters, with common choices being sigmoid, rectified linear unit function (RELU), hyperbolic tan function (TANH), and softmax. The number of hidden layers and the number of neurons in each of these layers are additional hyper-parameters of the model. The number of neurons in the input and the output layers are defined according to the classification or regression task at hand. Shallow neural networks are available in {\sc scikit-learn}\footnote{\url{https://scikit-learn.org/stable/modules/neural_networks_supervised.html}}.

Owing to their flexible structure and non-linearity, artificial neural networks are powerful algorithms, capable of describing extremely complex relations between the input data and the target variable. As just noted, these networks have many hyper-parameters, and they usually require a large amount of data to train on. Furthermore, due to their complexity, they tend to overfit the dataset, and various techniques, such as \emph{dropout}, are applied to overcome this issue. In addition, these networks are harder to interpret, compared to SVM or Random Forest. However, studies have shown that deep neural networks can greatly outperform traditional algorithms such as SVM, Random Forest, and shallow neural networks, given raw and complex data, such as images, spectra, and light-curves (see e.g., \citealt{huertas_company18, naul18, parks18} and references within).

\section{Unsupervised Learning}\label{sec:unsupervised_algs}

Unsupervised Learning is a general term that incorporates a large set of statistical tools, used to perform data exploration, such as clustering analysis, dimensionality reduction, visualization, and outlier detection. Such tools are particularly important in scientific research, since they can be used to make new discoveries or extract new knowledge from the dataset. For example, a cluster analysis that reveals two distinct clusters of planets might suggest that the two populations are formed through different formation channels, or, a successful embedding of a complex high-dimensional dataset onto two dimensions might suggest that the observed complexity can be attributed to a small number of physical parameters (e.g., the large variety of stellar spectra can be attributed to a single sequence in temperature, stellar mass, and luminosity; e.g., \citealt{hertzsprung09} and \citealt{russell14}). While visual inspection of the dataset can achieve these goals, it is usually limited to 3--12 dimensions (see lecture 2 by S. G. Djorgovski). Visual inspection becomes impractical with modern astronomical surveys, which provide hundreds to thousands of features per object. It is therefore necessary to use statistical tools for this task.

Unsupervised Learning algorithms take as an input only the measured features, without labels, and as such, they cannot be trained with some "ground truth". Their output is typically a non-linear and non-invertible transformation of the input dataset, consisting of an association of different objects to different clusters, low-dimensional representation of the objects in the sample, or a list of peculiar objects. Such algorithms consist of internal choices and a cost function, which does not necessarily coincide with our scientific motivation. For example, although K-means algorithm is often used to perform clustering analysis (see section \ref{sec:k_means}), it is not optimized to detect clusters, and it will reach a correct global optimum even if the dataset is not composed of clusters. In addition, these algorithms often have several \emph{external} free parameters, which cannot be optimized through the algorithm's cost function. Different choices of external parameters may result in significantly different outputs, resulting in different interpretations of the same dataset. The interpretation of the output of an unsupervised learning algorithm must be carried out with extreme caution, taking into account its internal choices and cost function, and the output's sensitivity to a change of external parameters. However, since unsupervised learning is typically used for exploration, not necessarily to reach a precisely determined goal, the lack of objective to optimize is often not critical. It does make it hard to know when one has really exhausted exploring all possibilities. 

\subsection{Distance Assignment}\label{sec:distances}

\begin{figure*}
	\centering
\includegraphics[width=0.98\textwidth]{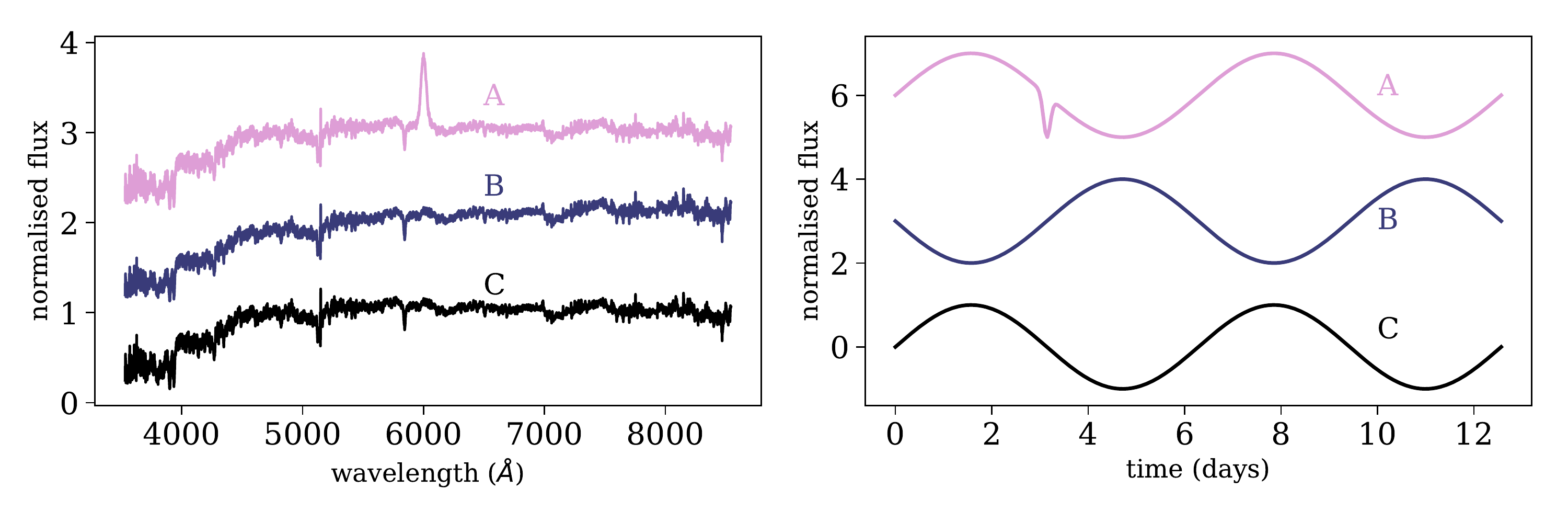}
\caption{Illustration of the disadvantages in using the Euclidean metric to measure distances between astronomical observations. The left panel shows three galaxy spectra, marked by \texttt{A}, \texttt{B}, \texttt{C}, where galaxy \texttt{A} appears visually different. However, the Euclidean distance between galaxy \texttt{B} and \texttt{C} is \emph{larger} than the distance between galaxy \texttt{A} and \texttt{B}, since most of the pixels in a galaxy spectrum are continuum pixels, and galaxy \texttt{B} is slightly bluer than galaxy \texttt{C}. The right panel shows three synthetic light-curves. Since the Euclidean metric is not invariant to horizontal shifts, the distance between light-curve \texttt{B} and \texttt{C} is larger than the distance to light-curve \texttt{A}, although the latter appears visually different.}
\label{f:distance_assignment_importance}
\end{figure*}

The first step in the large majority of unsupervised learning algorithms is distance assignment between the objects in the sample. In most of the cases it is assumed that the measured features occupy a euclidean space, and the distance between the objects in the sample is measured with euclidean metric. This choice might not be appropriate for some astronomical datasets, and using a metric that is more appropriate could improve the algorithm's performance. For example, when the input dataset consists of features that are extracted from astronomical observations, these features typically do not have the same physical units (e.g., stellar mass, temperature, size and morphology of a galaxy, bolometric luminosity, etc). In such cases, euclidean distance assignment will be dominated by features that are distributed over the largest dynamical range (e.g., black hole mass $\sim 10^{8}\,\mathrm{M_{\odot}}$), and will not be sensitive to the other features in the input dataset (e.g., stellar velocity dispersion $\sim 200\,\mathrm{km/sec}$). Therefore, when applying unsupervised machine learning algorithms to a set of derived features, it is advised to rescale the features (see also section \ref{sec:input_dataset}).

When applied to astronomical observations, such as spectra, light-curves, or images, the euclidean distance might not trace the distances a scientist would assign. This is partly since the euclidean metric implicitly assumes that all the features are equally important, which is not necessarily the case. Figure \ref{f:distance_assignment_importance} shows two examples of such scenarios. The left panel shows spectra of three galaxies, marked by \texttt{A}, \texttt{B}, and \texttt{C}, and the right panel shows three synthetic light-curves. In the galaxy case, galaxy \texttt{A} appears different, since it shows a strong emission line. Thus, we would expect that the distance between galaxy \texttt{A} and \texttt{B} will be larger than the distance between galaxy \texttt{B} and \texttt{C}. However, most of the pixels in a galaxy spectrum are continuum pixels, and the continuum of galaxy \texttt{B} is slightly bluer than the continuum of galaxy \texttt{C}. Since the spectra are dominated by continuum pixels, the euclidean distance between galaxy \texttt{B} and \texttt{C} will be \emph{larger} than the distance to galaxy \texttt{A}. That is, using a euclidean metric, galaxy \texttt{B} is different. Similarly in the light-curve case, since the euclidean metric is sensitive to horizontal shifts, the distance between light-curve \texttt{B} and \texttt{C} will be larger than the distance between light-curve \texttt{A} and \texttt{C}. In many scientific applications, we want to define a metric in which \texttt{A} stands out. Such metrics can be based on domain knowledge, where e.g., the metric is invariant to shifts, rotations, and flips, or they can be based on some measure of feature importance, e.g., emission line pixels being more important than continuum pixels. 

There are several distance measures which can be more appropriate for astronomical datasets and may result in improved performance, such as cross correlation-based distance (see e.g., \citealt{protopapas06, nun16, reis18b}), Distance Correlation, Dynamic Time Warp, Canberra Distance, and distance measures that are based on supervised learning algorithms (see \citealt{reis18b} for details). Section \ref{sec:unsup_forest} describes a general distance measure that is based on the Random Forest algorithm, and was shown to work particularly well on astronomical spectra. Since the Random Forest ranks features according to their importance, the Random Forest-based distance is heavily influenced by important features, and is less affected by irrelevant features.

\subsubsection{General Similarity Measure with Random Forest}\label{sec:unsup_forest}

So far, the discussion on Random Forest (section \ref{sec:ensemble_methods}) was focused on a supervised setting, where the model is trained to perform classification and regression. However, Random Forest can be used in an unsupervised setting, to estimate the distance between every pair of objects in the sample. \citet{shi06} presented such a method, and \citet{baron17}, followed by \citet{reis18a} and \citet{reis18b}, applied the method to astronomical datasets, with a few modifications that made the algorithm more suitable for such datasets. 

\begin{figure*}
	\centering
\includegraphics[width=0.49\textwidth]{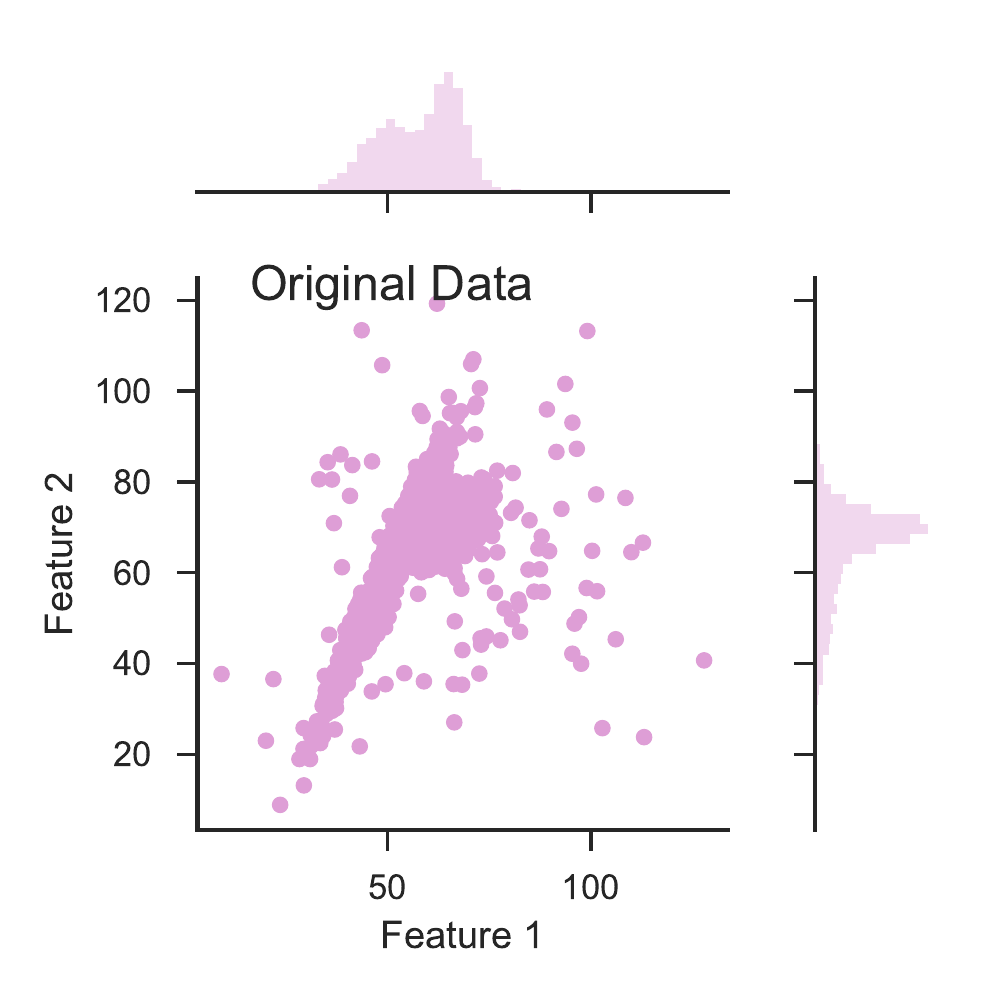}
\includegraphics[width=0.49\textwidth]{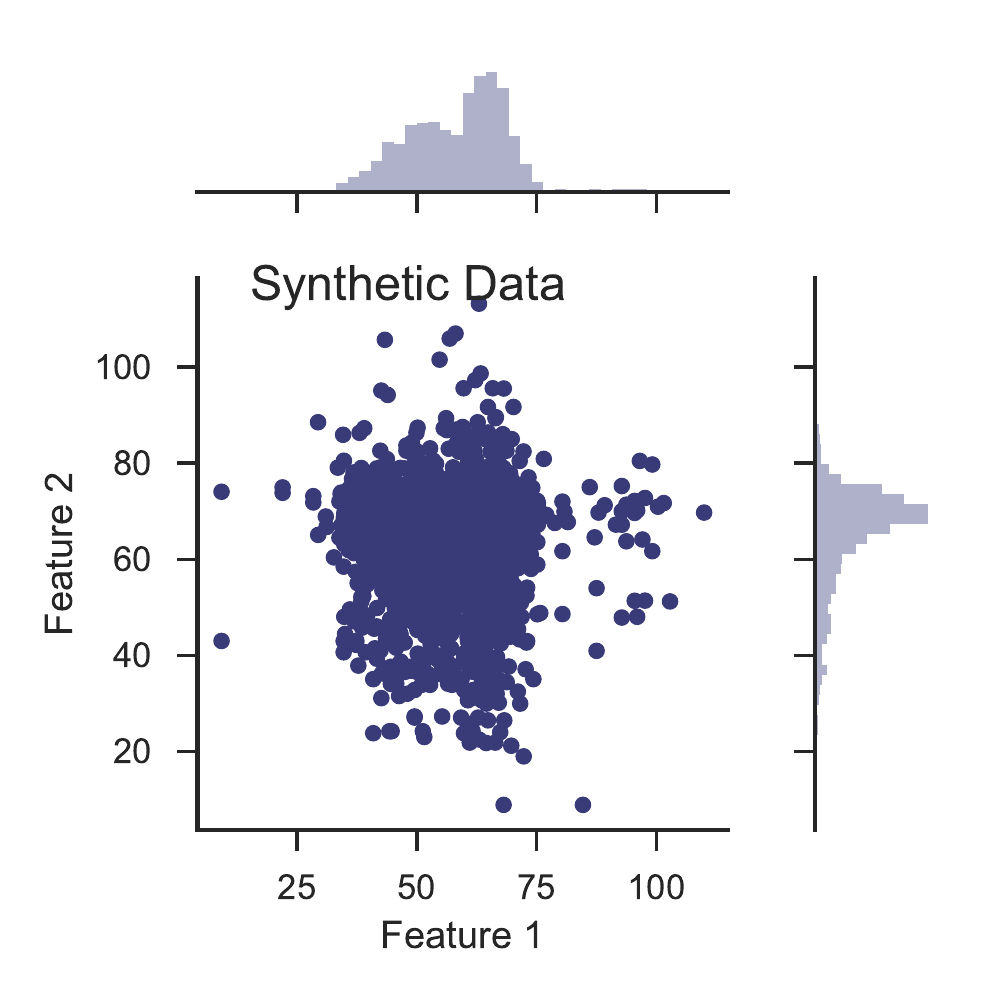}
\caption{Illustration of synthetic data construction, taken from \citet{baron17}, in a simplified example of a dataset with only two features. The left panel shows the distribution of the features of the original dataset, and their marginal distributions. The synthetic dataset (right panel) is constructed by sampling from the marginal distribution of each feature in the original dataset. The resulting dataset shows the same marginal distribution in its features, but is stripped of the covariance that was present in the original dataset.}
\label{f:rf_unsup}
\end{figure*}

In an unsupervised setting, the dataset consists only of measured features, without labels, and is represented by an $N \times M$ matrix, where every row represents an object ($N$ objects in the sample), and every column represents a feature ($M$ measured features per object). To translate the problem into a supervised learning problem that can be addressed with Random Forest, a \emph{synthetic} dataset is built. The synthetic dataset is represented by a $N \times M$ matrix, similarly to the original dataset, where each feature (column) in the synthetic data is built by sampling from the marginal distribution of the same feature in the original dataset. One can, instead, shuffle the values in every column of the original data matrix, resulting in a similar synthetic data matrix. The process of creating the synthetic data is illustrated in figure \ref{f:rf_unsup} with a simplified example taken from \citet{baron17}, where the objects in the dataset have only two features. The left panel shows the original dataset, with the marginal distributions in each of the features plotted on the top and on the right. The right panel shows the synthetic dataset, where the features show the same marginal distribution as in the original dataset, but stripped of the covariance seen in the original dataset. 

Once the synthetic dataset is constructed, the original dataset is labeled as class \texttt{1} and the synthetic dataset is labeled as class \texttt{2}, and Random Forest is trained to distinguish between the two classes. During this training phase, the Random Forest is trained to detect \emph{covariance}, since it is present only in the original dataset and not in the synthetic one. As a result, the most important features in the decision trees will be features that show correlations with others. Having the trained forest, the distance between different objects (in the original dataset) is defined as follows. Every pair of objects is propagated through all the decision trees in the forest, and their similarity is defined as the number of times they were both classified as class \texttt{1}, and reached the \emph{same} terminal leaf. This similarity $S$ can range between $0$ to the number of trees in our forest. $S$ is a measure of the similarity between these two objects since objects that have a similar path inside the decision tree have similar features, and as a consequence are represented by the same model (for more details see \citealt{baron17}). 

\citet{baron17}, \citet{reis18a}, and \citet{reis18b} showed that this definition of similarity between objects traces valuable information about their different physical properties. Specifically, they showed that such metric works particularly well for spectra, and traces information coming from different emission and absorption lines, and their connection to the continuum emission. \citet{reis18a} applied this method to infrared spectra of stars, and showed that this metric traces physical properties of the stars, such as metallicity, temperature, and surface gravity. The algorithm was implemented in {\sc python} and is publicly available on {\sc github}\footnote{\url{https://github.com/dalya/WeirdestGalaxies}}.

\subsection{Clustering Algorithms}\label{sec:clustering}

Clustering analysis, or clustering, is the task of grouping objects in the sample, such that objects in the same group, which is called a cluster, are more similar to each other than to objects in other groups. The definition of a cluster changes from one algorithm to the next, and this section describes centroid-based clustering (K-means; section \ref{sec:k_means}) and connectivity-based clustering (Hierarchical clustering; section \ref{sec:hiararchical}). Also popular is distribution-based clustering, with the prominent method being Gaussian Mixture Models, which I do not discuss in this document (see e.g., \citealt{de_souza17} and references within). Throughout this section, I give examples with a simplified dataset that consists of two features. Obviously, a two-dimensional dataset can be visualized and clusters can be detected manually, however, these algorithms can be used to search for clusters in complex high-dimensional datasets.

\subsubsection{K-means}\label{sec:k_means}

\begin{figure*}
	\centering
\includegraphics[width=0.98\textwidth]{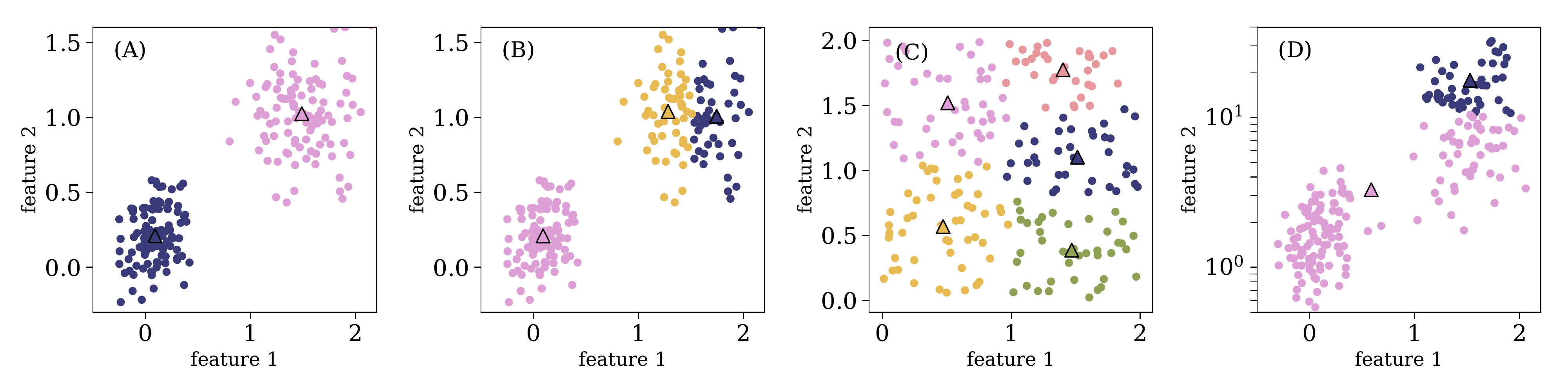}
\caption{Four examples of K-means application to different two-dimensional datasets, where the objects in the sample are marked with circles and the clusters centroids are marked with triangles. The different colors represent different clusters. Panel (A) shows K-means application with $k=2$ and a euclidean distance. Panel (B) shows K-means application to a similar dataset, but with $k=3$. Panel (C) shows K-means output for a dataset without clear clusters, and panel (D) shows the result for a dataset with features which are distributed over different dynamical scales. }
\label{f:k_means_examples}
\end{figure*}

One of the most widely used clustering methods is K-means, which is a centroid-based clustering algorithm (e.g., \citealt{macqueen67}). K-means is simple and robust, even when performing clustering analysis in a high-dimensional space. It was used in Astronomy in various contexts, to study stellar and galaxy spectra, X-ray spectra, solar polarization spectra, spectra from asteroids, and more (see e.g., \citealt{balazs96, hojnacki07, galluccio08, almeida10, simpson12, almeida13, garcia_dias18} and references therein).

The first step of K-means is distance assignment between the objects in the sample. The default distance is the euclidean metric, but other metrics, which are more appropriate for the particular dataset at hand, can be used. Then, the algorithm selects $k$ random objects from the dataset which serve as the initial centroids, where $k$ is an external free parameter. Each object in the dataset is then assigned to its closest of the $k$ centroids. Then, new cluster centroids are computed by taking the average position of the objects that are associated with the given cluster. These two steps, re-assigning objects to a cluster according to their distance from the centroid and recomputing the cluster centroids, are repeated iteratively until reaching convergence. Convergence can be defined in several manners, for example, when the large majority of the objects are no longer reassigned to different centroids (90\% and more), or when the cluster centroids converge to a set location. The output of the algorithm consists of the cluster centroids, and an association of the different objects to the different clusters. K-means is available in the {\sc scikit-learn} library\footnote{\url{https://scikit-learn.org/stable/modules/generated/sklearn.cluster.KMeans.html}}.

Panel (A) in Figure \ref{f:k_means_examples} shows an application of K-means to a two-dimensional dataset, setting $k=2$ and using euclidean distances, where the dots represent the objects in the sample, triangles represent the cluster centroids, and different colors represent different clusters. As noted in the previous paragraph, K-means starts by randomly selecting $k$ objects as the initial cluster centroids. In some cases, different random assignments of initial centroids might result in different outputs, particularly when K-means converges to a local minimum. To avoid it, one can run K-means several times, each time with different randomly-selected centroids, and select the output that results in the minimum sum of squared distances between the objects and their centroids. Panel (B) in Figure \ref{f:k_means_examples} shows an application of K-means to a similar dataset, but with $k=3$. $k$ is an example of an external parameter that cannot be optimized with the cost function, since the cost function decreases monotonically with $k$. Obviously, setting $k$ to be equal to the number of objects in the sample will result in the minimum possible cost of zero, since each object will be defined as its own cluster. Finding the best $k$ is not trivial in most of the cases, and studies either use the \emph{elbow method}, or define probability-based scores to constrain it (see e.g., \citealt{almeida10}). In some cases, the distribution of the distances of all the objects in the sample contains several distinguishable peaks, and can guide the selection of the correct $k$.

Panel (C) in Figure \ref{f:k_means_examples} shows an application of K-means to a dataset with no clear clusters. This is an example in which the output consists of clusters while the dataset does not show clear clusters. To test for such cases, one can compare the distribution of distances between objects within the same cluster to the typical distance between cluster centroids. In this particular example, these are roughly similar, suggesting that there are no clear clusters in the dataset. Panel (D) shows an application of K-means to a dataset with features that are distributed over different dynamical scales, and one can see that K-means failed in finding the correct clusters. The K-means cost function is based on the summed distances between the objects and their centroids, and since the distances between the objects are much larger in feature 2 (y-axis), the optimal output is completely dominated by its values. To avoid such issues, it is recommended to either normalize the features, or rank-order them before applying K-means. K-means will also fail when the dataset contains outliers, since they can have a significant impact on the centroids placements, and therefore on the resulting clusters. Thus, outliers should be removed before applying K-means to the dataset.

\subsubsection{Hierarchical Clustering}\label{sec:hiararchical}

Hierarchical clustering is another popular algorithm in cluster analysis, aiming at building a hierarchy of clusters (e.g., \citealt{ward63}). The two main types of Hierarchical clustering are Agglomerative Hierarchical clustering, also named the "bottom-up" approach, where each object starts as an individual cluster and clusters are merged iteratively, and Divisive Hierarchical clustering, or "top-down", where all the objects start in one cluster, then split recursively into smaller clusters. Hierarchical clustering has been applied to various astronomical datasets, such as X-ray spectra, extracted features from galaxy images, and absorption spectra of interstellar gas (see e.g., \citealt{hojnacki07, baron15, hocking15, peth16, ma18}). The discussion in this section is focused on Agglomerative Hierarchical clustering. 

The first step of Hierarchical clustering is also assigning distances between the objects in the sample, using the euclidean metric by default. All the objects in the sample start as one-sized clusters. Then, the algorithm merges the two closest clusters into a single cluster. The process is repeated iteratively, merging the two closest clusters into a single cluster, until the dataset consists of a single cluster, which includes many smaller clusters. To do so, one must define a distance between clusters that contain more than one object, also referred to as "linkage method". There are several different linkage methods, which include "complete linkage", where the distance between the clusters is defined as the \emph{maximal} distance between the objects in the two clusters, "single linkage", where the distance between the clusters is defined as the \emph{minimal} distance between the objects in the clusters, "average linkage", where the distance is defined as the \emph{average} distance between the objects in the clusters, and "ward linkage", which minimizes the variance of the clusters merged. The linkage method is an external free parameter of the algorithm, and it has a significant influence on the output.

\begin{figure}
\includegraphics[width=0.49\textwidth]{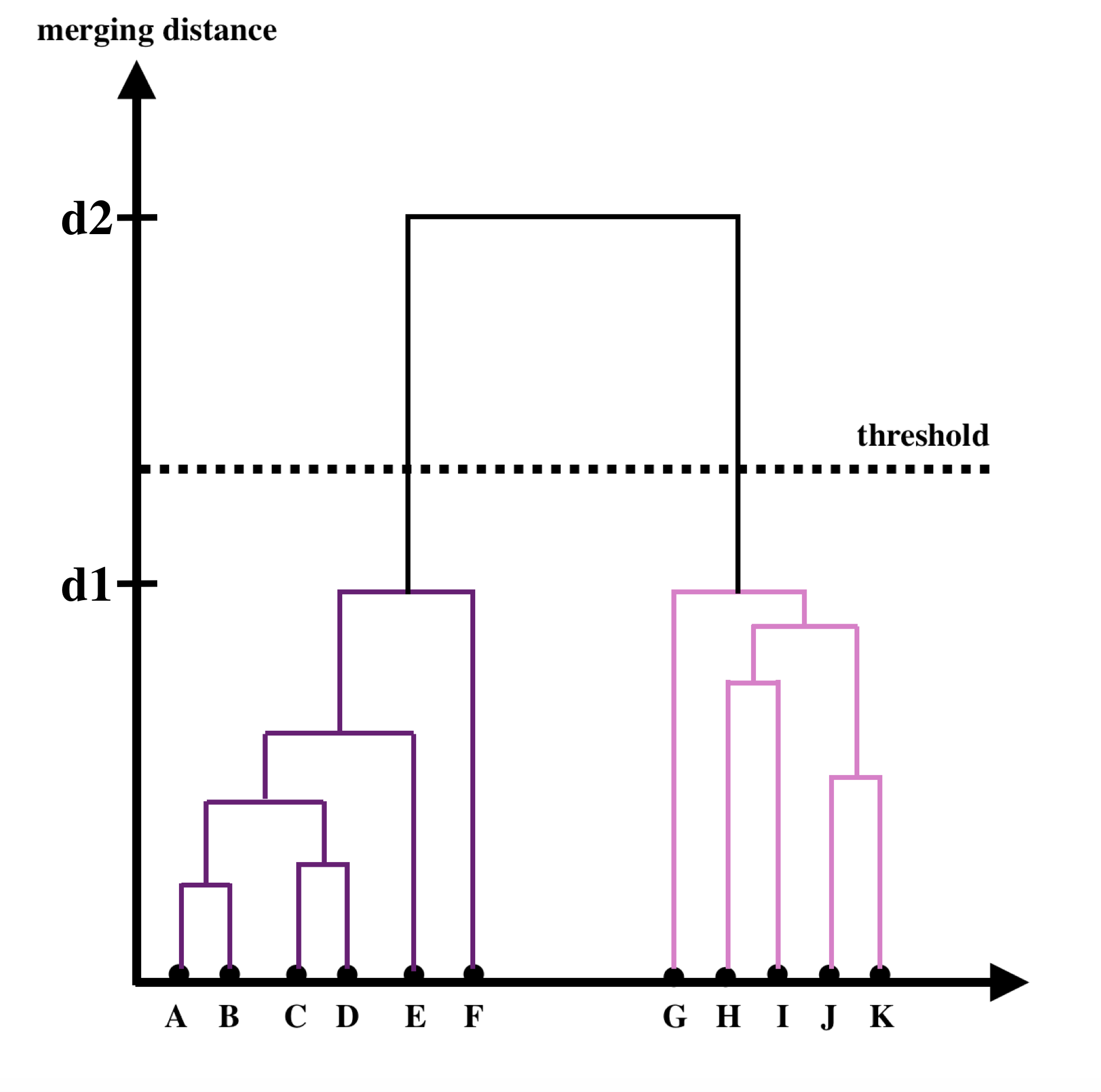}
\caption{Visualization of a dendrogram for a dataset with 11 objects, marked by \texttt{A}, \texttt{B}, .., \texttt{K}. The dendrogram represents the hierarchical merging history, where the y-axis represents the distance at which two clusters were merged. The clusters that merged with a shorter distance were merged earlier in the process, and in this case, the merger history is \texttt{A} and \texttt{B}, \texttt{C} and \texttt{D}, \texttt{(A,B)} and \texttt{(C,D)}, \texttt{J} and \texttt{K}, and so on. The dashed horizontal line represents the threshold $t$ that is used to define the final clusters in the dataset (see text for more details). }
\label{f:dendrogram}
\end{figure}

\begin{figure*}
	\centering
\includegraphics[width=0.98\textwidth]{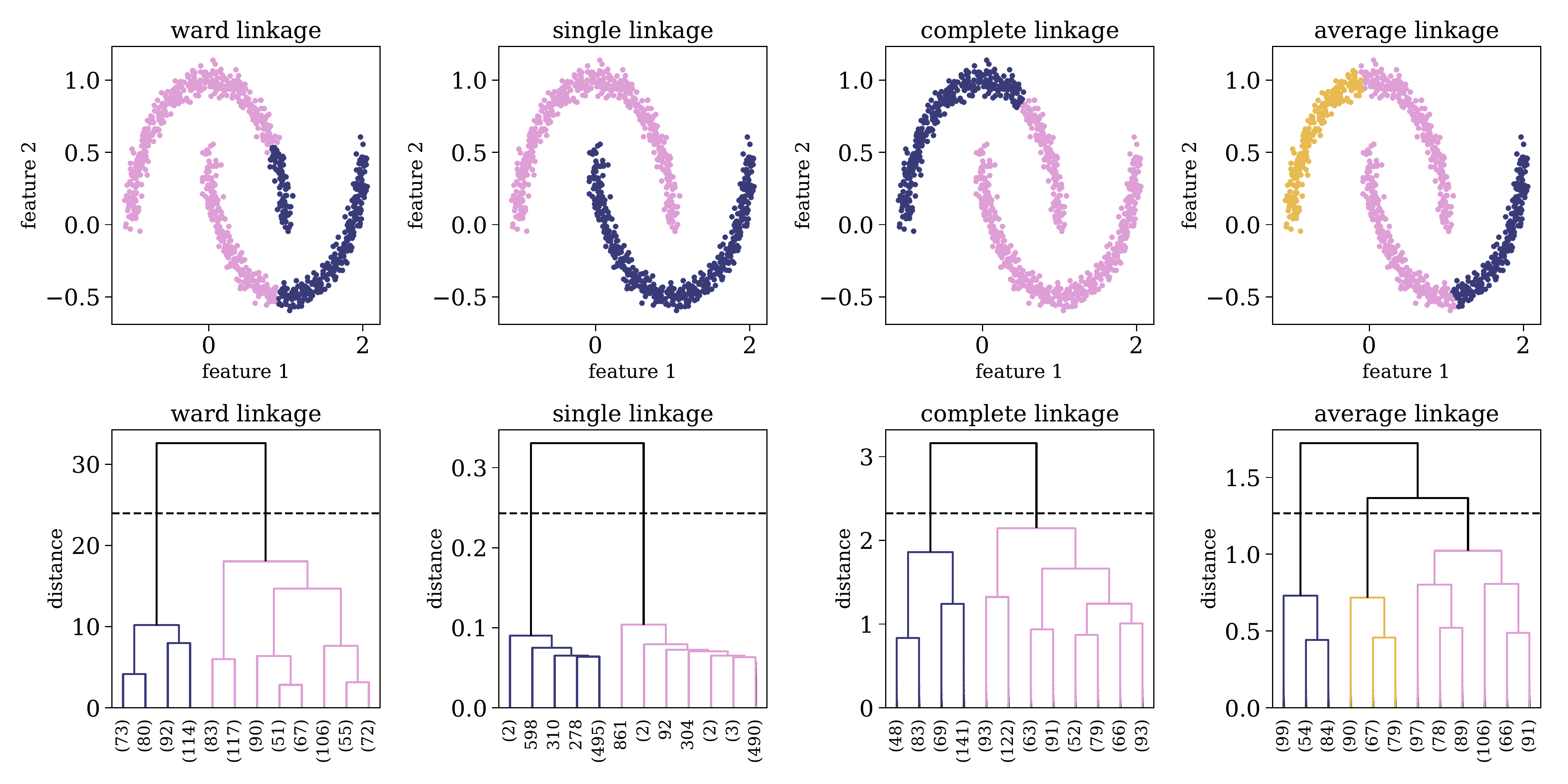}
\caption{Application of Hierarchical clustering to a two-dimensional dataset, using different linkage methods, and setting $t$ to be 0.7 of the maximal merging distance. The top panels show the distribution of the objects in the dataset, colored by the final cluster association. The bottom panels show the resulting dendrograms of each linkage method, and the dashed horizontal line represents the threshold $t$ used to define the final clusters. The dendrograms are truncated for representational purposes, and the number objects in each truncated branch is indicated on the x-axis in parentheses.}
\label{f:hierarchical_clustering_example}
\end{figure*}

The result of Hierarchical clustering is usually visualized with a dendrogram. Figure \ref{f:dendrogram} shows an application of Hierarchical clustering to a dataset with 11 objects, marked by \texttt{A}, \texttt{B}, .., \texttt{K} in the diagram. The dendrogram represents the history of the hierarchal merging process, with the vertical axis showing the distance at which two clusters were merged. Clusters \texttt{A} and \texttt{B} were merged first, since their merging distance is the shortest (up to this point the clusters contain a single object). Then, clusters \texttt{C} and \texttt{D} were merged. Following that, clusters \texttt{(A, B)} and \texttt{(C, D)} were merged, since they were the clusters with the shortest distance. The next two merging clusters are \texttt{J} and \texttt{K}, and the process continues until clusters \texttt{(A, B, C, D, E, F)} and \texttt{(G, H, I, J, K)} are merged into a single cluster. The dendrogram can be used to study the structure of the dataset, in particular, it can be used to infer the number of clusters in the dataset. The example in Figure \ref{f:dendrogram} suggests that the dataset consists of two clusters, \texttt{(A, B, C, D, E, F)} and \texttt{(G, H, I, J, K)}, since the merging distance of inter-cluster objects ($d1$ in the figure) is much shorter than the merging distance of the two final clusters ($d2$ in the figure). The dashed horizontal line represents the threshold $t$ used in Hierarchical clustering for the final cluster definition. Groups that are formed beneath the threshold $t$ are defined as the final clusters in the dataset, and are the output of the algorithm. $t$ is an external free parameter of the algorithm, and cannot be optimized using the cost function. It has a significant effect on the resulting clusters, in particular, as $t$ decreases, the number of resulting clusters increases. In some cases, the dendrogram can be used to estimate a "natural" threshold value. Agglomerative Hierarchical clustering is available in {\sc scikit-learn}\footnote{\url{https://scikit-learn.org/stable/modules/generated/sklearn.cluster.AgglomerativeClustering.html}}. Visualization of the resulting dendrogram can be done with the {\sc scipy} library, and is presented in the hands-on tutorials\footnote{\url{https://github.com/dalya/IAC_Winter_School_2018}; see also: \\ \url{https://joernhees.de/blog/2015/08/26/scipy-hierarchical-clustering-and-dendrogram-tutorial/}}. 

\begin{figure*}
	\centering
\includegraphics[width=0.98\textwidth]{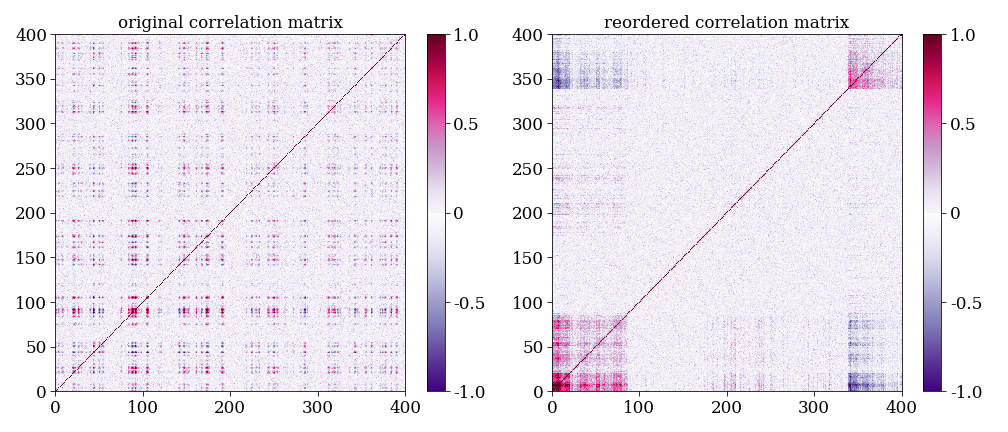}
\caption{Application of Hierarchical clustering to reorder and visualize a complex correlation matrix. The left panel shows a correlation matrix calculated for a dataset with 400 objects, where the colorbar represents the Pearson correlation coefficient between every pair of objects. The rows and columns are ordered according to the object index, and little information can be extracted from such a visualization. The right panel shows the same correlation matrix, reordered according to the appearance order of the objects in the dendrogram. The latter representation reveals interesting structures, and can be used to interpret the dataset.}
\label{f:hierarchical_clustering_reordering_example}
\end{figure*}

Figure \ref{f:hierarchical_clustering_example} shows an example of Hierarchical clustering application to a two-dimensional dataset, using different linkage methods, setting $t$ to be 0.7 of the maximal merging distance. The top panels show the distribution of the objects in the dataset, colored by the final clusters detected by the algorithm, and the bottom panels show the dendrograms for each of the linkage methods. Looking at the first row, one can see that different linkage methods result in different cluster definitions, in particular, both the number of detected clusters and the association of objects to clusters change for different linkage methods. In this particular example, it is clear that the single linkage returns the correct output, however, for high-dimensional datasets such a visualization is not possible. To select the "correct" linkage method for the particular dataset at hand, it is advised to examine the resulting dendrograms. In this specific case, the second row shows that the single linkage-based dendrogram reveals the most significant clustering, with two detected clusters. A general rule of thumb to select of the "correct" linkage method is by selecting the linkage that results in the largest difference between the merging distance of the resulting clusters and the final, maximal, merging distance. 

As shown in figure \ref{f:hierarchical_clustering_example}, Hierarchical clustering can be used to detect clusters which other clustering algorithms, such as K-means and Gaussian Mixture Models, cannot detect. Since it is based on connectivity, it can be used to detect clusters that are distributed over a non-trivial manifold (this is usually possible only with the single linkage method). Furthermore, Hierarchal clustering is less sensitive to outliers in the dataset, since these will be merged last, and thus will not effect the structure of the dendrogram and the resulting clusters that merged before that. Perhaps the most interesting application of Hierarchical clustering is reordering and visualizing complex distance or correlation matrices. The left panel of Figure \ref{f:hierarchical_clustering_reordering_example} shows as example of a correlation matrix, calculated for a complex dataset with 400 objects. The rows and columns of the correlation matrix are ordered according to the object index, and clearly, this representation conveys little information about the structure of the dataset. Instead, one can perform Hierarchical clustering and extract a dendrogram for this particular dataset. Then, one can rearrange the objects in the correlation matrix according to their order of appearance in the dendrogram. The reordered correlation matrix is shown in the right panel of Figure \ref{f:hierarchical_clustering_reordering_example}, where one can find at least two clusters of objects, such that objects that belong to a given cluster show strong correlations, and objects that belong to different clusters show a strong negative correlation. Therefore, the process described above may reveal rich structures in the dataset, which may allow one to explore and extract information from it, even without performing cluster analysis (see also \citealt{de_souza15}). 

\subsection{Dimensionality Reduction Algorithms}\label{sec:dim_red}

Dimensionality reduction refers to the process of reducing the number of features in the original dataset, either by selecting a subset of the features that best describe the dataset, or by constructing a new set of features that provide a good description of the dataset. Some of the dimensionality reduction algorithms provide principle components or prototypes, which are a small set of objects that have the same dimensions as the objects in the original dataset, and are used to represent all the objects in the sample. Other algorithms aim at embedding the high-dimensional dataset onto a low-dimensional space, without using principle components or prototypes. When applying dimensionality reduction algorithms to data, we always lose some information. Our goal is to choose an algorithm that retains most of the \emph{relevant} information, where relevant information strongly depends on our scientific motivation.

Dimensionality reduction is useful for a variety of tasks. In a supervised learning setting, since many algorithms are not capable of managing thousands of features, dimensionality reduction is used to decrease the number of features under consideration, by removing redundancy in the original set of features. Although not very popular in Astronomy, dimensionality reduction is also used for compression, and will become more relevant for surveys such as the SKA \citep{dewdney09}, where keeping all the data is no longer possible. Perhaps most importantly, dimensionality reduction can be used to visualize and interpret complex high-dimensional datasets, with the goal of uncovering hidden trends and patterns. 

\begin{figure}
\includegraphics[width=0.49\textwidth]{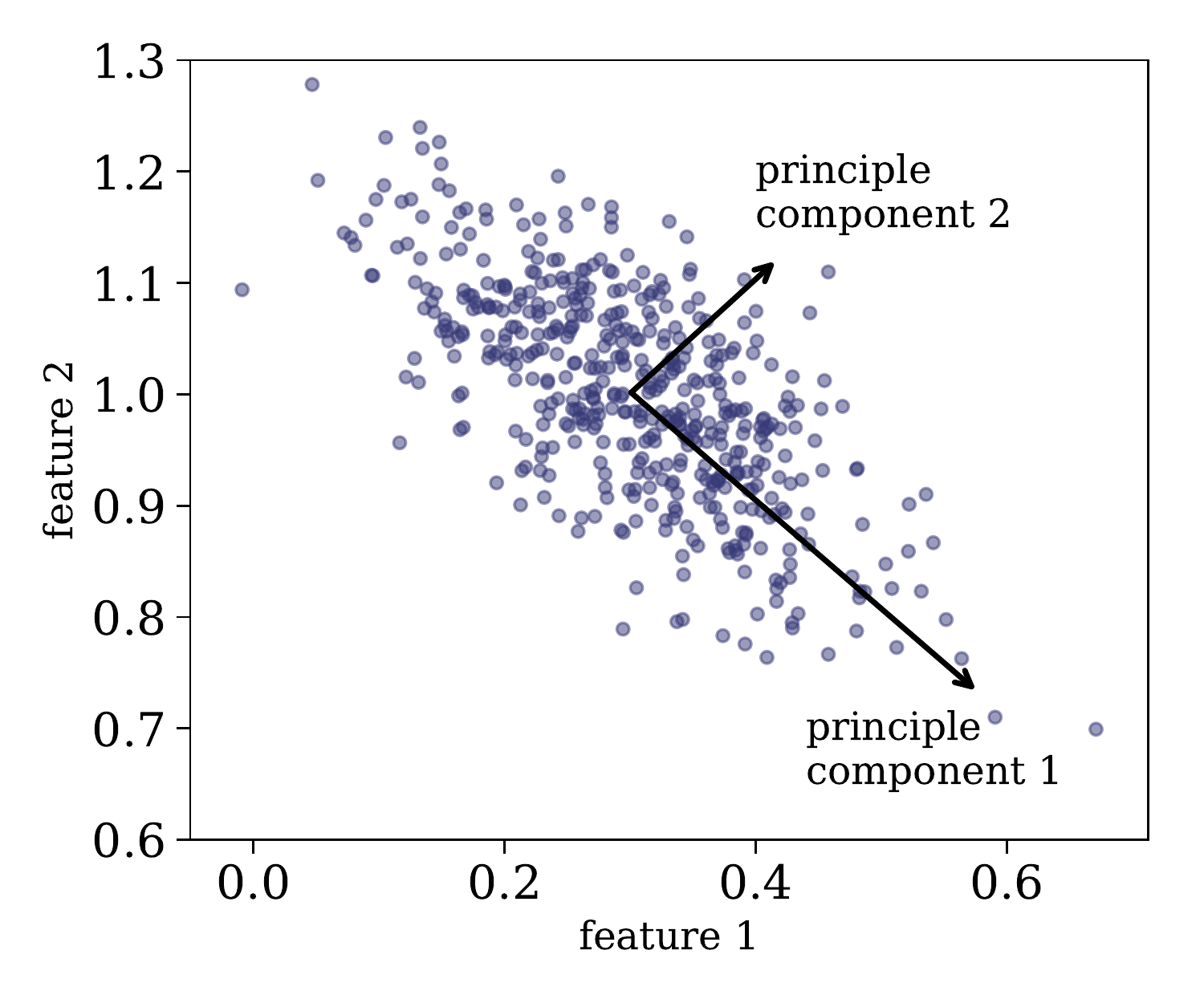}
\caption{Application of PCA to two-dimensional dataset. The two principle components are marked with black arrows, where principle component 1 accounts for most of the variance in the data, and principle component 2, which is orthogonal to principle component 1, accounts for the rest of the variance in the data. Every object in the sample can be accurately represented as a linear combination of the two principle components, and can be represented approximately using the first principle component. }
\label{f:pca_example_2d}
\end{figure}

\begin{figure*}
	\centering
\includegraphics[width=0.6\textwidth]{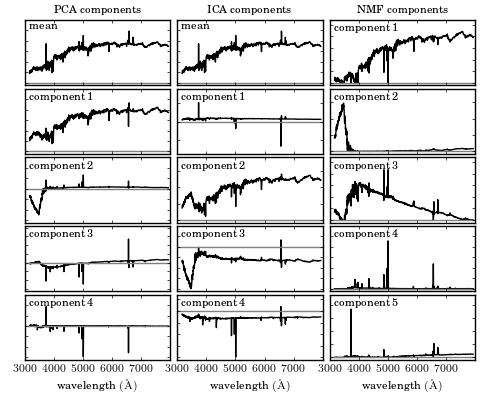}
\caption{Application of PCA, ICA, and NMF to a sample of SDSS spectra, taken from \citet{vanderplas12}. Additional details on ICA and NMF can be found in \citet{ivezic14}. The columns represent different algorithms, and the rows represent the resulting components using each of the algorithms, sorted by their importance. The grey horizontal line represents zero flux. While the first two PCA components resemble galaxy spectra (with an old stellar population), the next three components do not represent a physical component in galaxy spectra, in particular, they show negative flux values. On the other hand, the NMF components resemble more physical components, with the first corresponding to an old stellar population, the second corresponding to blue continuum emission, which might be due to an AGN or O- and B-type stars, the third corresponding to younger stellar population (A-type stars), and the forth and fifth components corresponding to emission lines.}
\label{f:pca_ica_nmf_example}
\end{figure*}

\subsubsection{Principle Component Analysis}\label{sec:pca}

Principle Component Analysis (PCA) is a linear feature projection, which transforms data in high-dimensional space to a lower-dimensional space, such that the variance of the dataset in the low-dimensional representation is maximized. In practice, PCA constructs a covariance matrix of the dataset and computes its eigenvectors. The eigenvectors that correspond to the largest eigenvalues are used to reconstruct a large fraction of the variance in the original dataset. These eigenvectors, also called principle components, are arranged such that the first principle component has the largest possible variance, and each succeeding component has the highest possible variance, under the constraint that it is orthogonal to the preceding components. The number of principal components is at most the number of features in the dataset. Every object in the sample can be represented by a linear combination of the principle components, where the representation is accurate only when \emph{all} the principle components are used. When a subset of the principle components is used, the representation is approximate, resulting in a dimensionality reduction. The coefficients of the linear combination of principle components can be used to embed the high-dimensional dataset onto a low-dimensional plane (typically 2 or 3 dimensions). Figure \ref{f:pca_example_2d} shows an application of PCA to a two-dimensional dataset, where the two principle components are marked with black arrows. One can see that the first principle component is oriented towards the direction with the maximal variance in the data, and the second principle component, which is orthogonal to the first one, describes the remaining variance.

PCA is among the most popular tools in Astronomy, and it has been used to search for multivariate correlations in high-dimensional datasets, estimate physical parameters of systems from their spectra, decompose complex spectra into a set of principle components which are then used as empirical templates, and more (e.g., \citealt{boroson92, djorgovski95, zhang06, vandenberk06, rogers07, fiorentin07, bailey12}). It is simple to use, has no free parameters, and is easily interpretable. However, PCA performs a \emph{linear} decomposition of the objects in the sample, which is not appropriate in many contexts. For example, absorption lines and dust extinction are multiplicative effects which cannot be described by a linear decomposition. Furthermore, PCA tends to find linear correlations between variables, even if those are non-linear, and it fails in cases where the mean and the covariance are not enough to define the dataset. Since it constructs its principle components to trace the maximal variance in the data, it is extremely sensitive to outliers, and these should be removed prior to applying PCA. Finally, the principle components of the dataset can contain negative values, which is also not appropriate in many astronomical setups. For example, applying PCA to galaxy spectra results in principle components with negative values, which is of course not physical, since the emission of a galaxy is a sum of \emph{positive} contributions of different light sources, \emph{attenuated} by absorbing sources such as dust or gas.

Finally, it is worth noting two additional techniques, Independent Component Analysis (ICA; \citealt{hyvarinen00}) and Non-Negative Matrix Factorization (NMF; \citealt{paatero94}), which are useful for a variety of tasks. ICA is a method used to separate a multivariate signal into additive components, which are assumed to be non-Gaussian and statistically independent from each other. ICA is a very powerful technique, which is often invoked in the context of blind signal separation, such as the "cocktail party problem". NMF decomposes a matrix into the product of two non-negative ones, and is used in Astronomy to decompose observations to non-negative components. Figure \ref{f:pca_ica_nmf_example} shows an application of PCA, ICA, and NMF, taken from \citet{vanderplas12}, on SDSS spectra (see \citealt{ivezic14} for additional details about ICA and NMF). The columns represent different algorithms, and the rows represent the resulting components using each of the algorithms, sorted by their importance. One can see that while the first two PCA components resemble galaxy spectra (with an old stellar population), the next three components do not represent a physical component in galaxy spectra, in particular, they show negative flux values. On the other hand, the NMF components resemble more physical components, with the first corresponding to an old stellar population, the second corresponding to blue continuum emission, which might be due to an AGN or O- and B-type stars, the third corresponding to younger stellar population (A-type stars), and the forth and fifth components corresponding to emission lines. Obviously, the resemblance is not perfect and one can see residual emission line components in the first and third components (in the first component these are described by absorption lines with central wavelengths corresponding to the strongest emission lines in galaxy spectra). The three algorithms are available in {\sc scikit-learn}\footnote{\url{https://scikit-learn.org/stable/modules/generated/sklearn.decomposition.PCA.html}\\ \url{https://scikit-learn.org/stable/modules/generated/sklearn.decomposition.FastICA.html}\\ \url{https://scikit-learn.org/stable/modules/generated/sklearn.decomposition.NMF.html}}.

\subsubsection{t-Distributed Stochastic Neighbor Embedding}\label{sec:tsne}

t-Distributed Stochastic Neighbor Embedding (tSNE; \citealt{vanDerMaaten08}) is a non-linear dimensionality reduction technique that embeds high-dimensional data in a low dimensional space, typically of two or three dimensions, and is mostly used to visualize complex datasets. The algorithm models every high-dimensional object using a two (or three) dimensional point, such that similar objects are represented by nearby points, whereas dissimilar objects are represented by distant points, with a high probability. tSNE has been used in Astronomy to visualize complex datasets and distance matrices, and study their structure (e.g., \citealt{lochner16, anders18, nakoneczny18, reis18a, alibert19, giles19, moreno19}).

\begin{figure*}
	\centering
\includegraphics[width=0.98\textwidth]{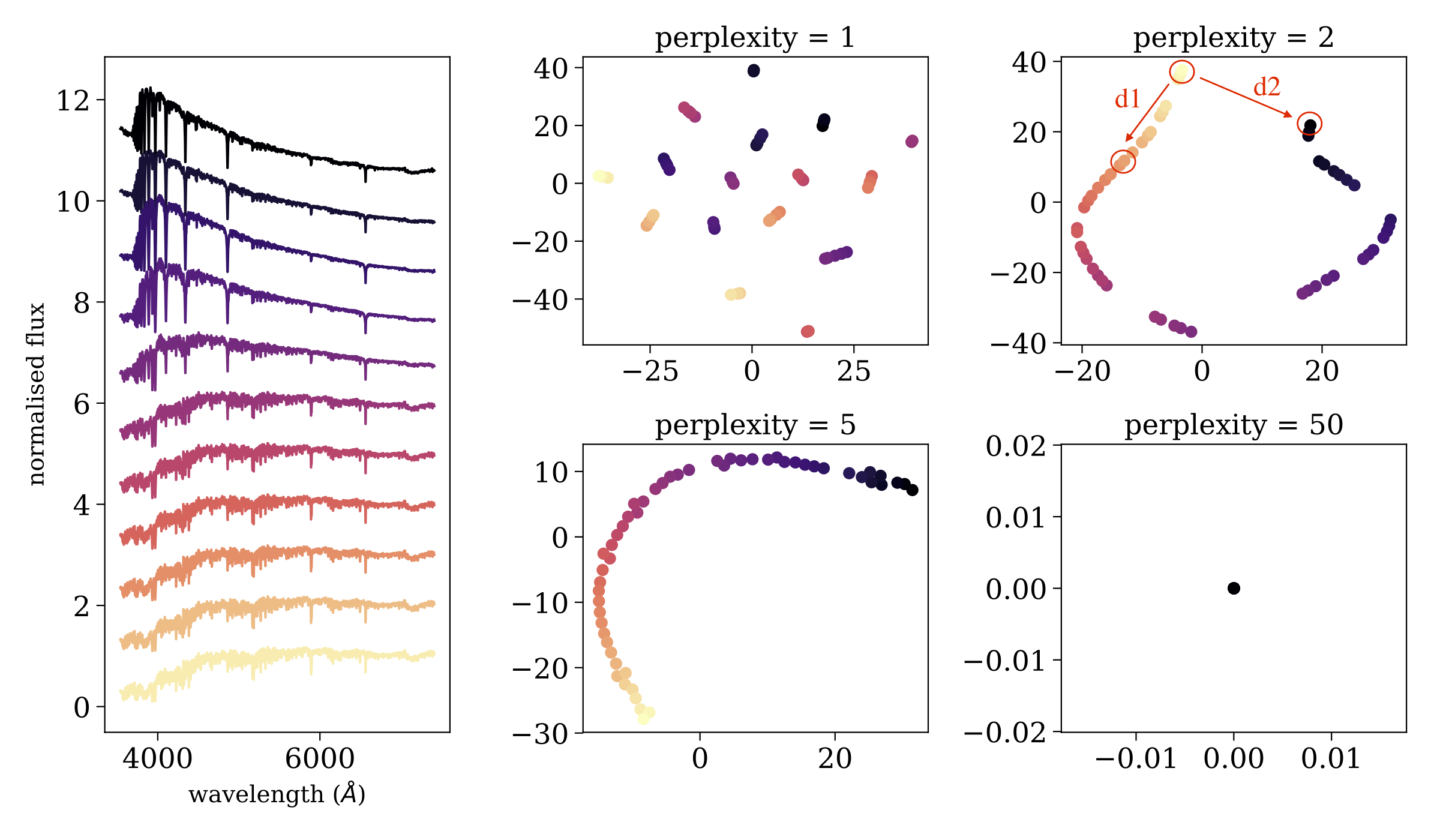}
\caption{Application of tSNE to a sample of 53 synthetic single stellar population models, with different ages. The left panel shows 11 out of the 53 spectra, where each spectrum consist of 4\,300 flux values. The right panels show the resulting two-dimensional embedding using different perplexity values, where every point in the 2D plane is colored according to the stellar age, such that purple points represent young stellar populations, and yellow points represent old stellar populations.}
\label{f:tsne_example}
\end{figure*}

The first step of tSNE is to assign distances between the objects in the sample, using the euclidean metric by default. Then, tSNE constructs a probability distribution over pairs of high-dimensional objects such that similar objects have a high probability of being picked, whereas dissimilar objects have an extremely low probability of being picked. This is achieved by modeling the probability distribution using a Gaussian kernel, which depends on the assigned distance between the objects, and a scale parameter, named the \emph{perplexity}, which is an external free parameter of the algorithm. The perplexity affects the neighborhood of objects being considered, in terms of their probability of being selected, where a small perplexity results in a very small neighborhood around a given object, and a large perplexity results in a larger neighborhood. The algorithm then embeds the high-dimensional objects into a low dimensional space, such that the probability distribution over pairs of points in the low-dimensional plane will be as similar as possible to the probability distribution in the high-dimensional space. The axes in the low-dimensional representation are meaningless and not interpretable. tSNE is available in the {\sc scikit-learn} library\footnote{\url{https://scikit-learn.org/stable/modules/generated/sklearn.manifold.TSNE.html}}.

Figure \ref{f:tsne_example} shows an application of tSNE to a sample of 53 single stellar population models, taken from the MILES library \citep{vazdekis10}. The left panel shows 11 out of the 53 spectra, where each spectrum has 4\,300 flux values, and therefore 4\,300 features, ordered by age. While the dataset appears complex, it actually represents a one-dimensional manifold, where all the observed properties can be attributed to a change in a single parameter, the age. Therefore, we expect that in the low dimensional representation, all the objects will occupy a single line. The right panels show the tSNE two-dimensional embedding for different perplexity values. When the perplexity is set to $p=1$, the neighborhoods considered by tSNE are too small, and the data is represented by many small clusters, although the dataset does not consist of clusters. When the perplexity is set to $p=2$ or $p=5$, the dataset is represented by an almost perfect one-dimensional manifold, tracing the correct structure of the data. When the perplexity is too high, e.g. $p=50$, all the points are plotted at the origin, and no structure can be seen. The perplexity has a significant effect on the resulting embedding, and cannot be optimized using the tSNE cost function. It represents the scale of the clusters tSNE is sensitive to, and setting a small perplexity value allows detection of small clusters, while setting a large perplexity value allows study of the global structure of the dataset. However, one must take into account that the low-dimensional embeddings by tSNE sometimes show small clusters, even if these do not exist in the dataset. In addition, one must be careful when trying to interpret long distances in the tSNE map. For example, when $p=2$, tSNE embeds the objects considering only neighborhoods of up to about the 6 nearest neighbors around each point, thus, distances smaller than those spanned by this neighborhood are conserved, while longer distances are not. While the distances $d1$ and $d2$ appear similar in the tSNE map in Figure \ref{f:tsne_example}, the distances between the corresponding objects in the original distance matrix are different by a factor of more than five.

Finally, it is worth noting an additional tool, Uniform Manifold Approximation and Projection (UMAP; \citealt{mcInnes18}), which can be used to perform non-linear dimensionality reduction. UMAP is a fast algorithm, that supports a wide variety of distance metrics, including non-metric distance functions such as cosine distance and correlation distance. Furthermore, \citet{mcInnes18} show that UMAP often outperforms tSNE at preserving global structures in the input dataset. UMAP is implemented in {\sc python} and is publicly available on {\sc github}\footnote{\url{https://github.com/lmcinnes/umap}}.

\subsubsection{Autoencoders}\label{sec:auto_encoders}

\begin{figure}
\includegraphics[width=0.5\textwidth]{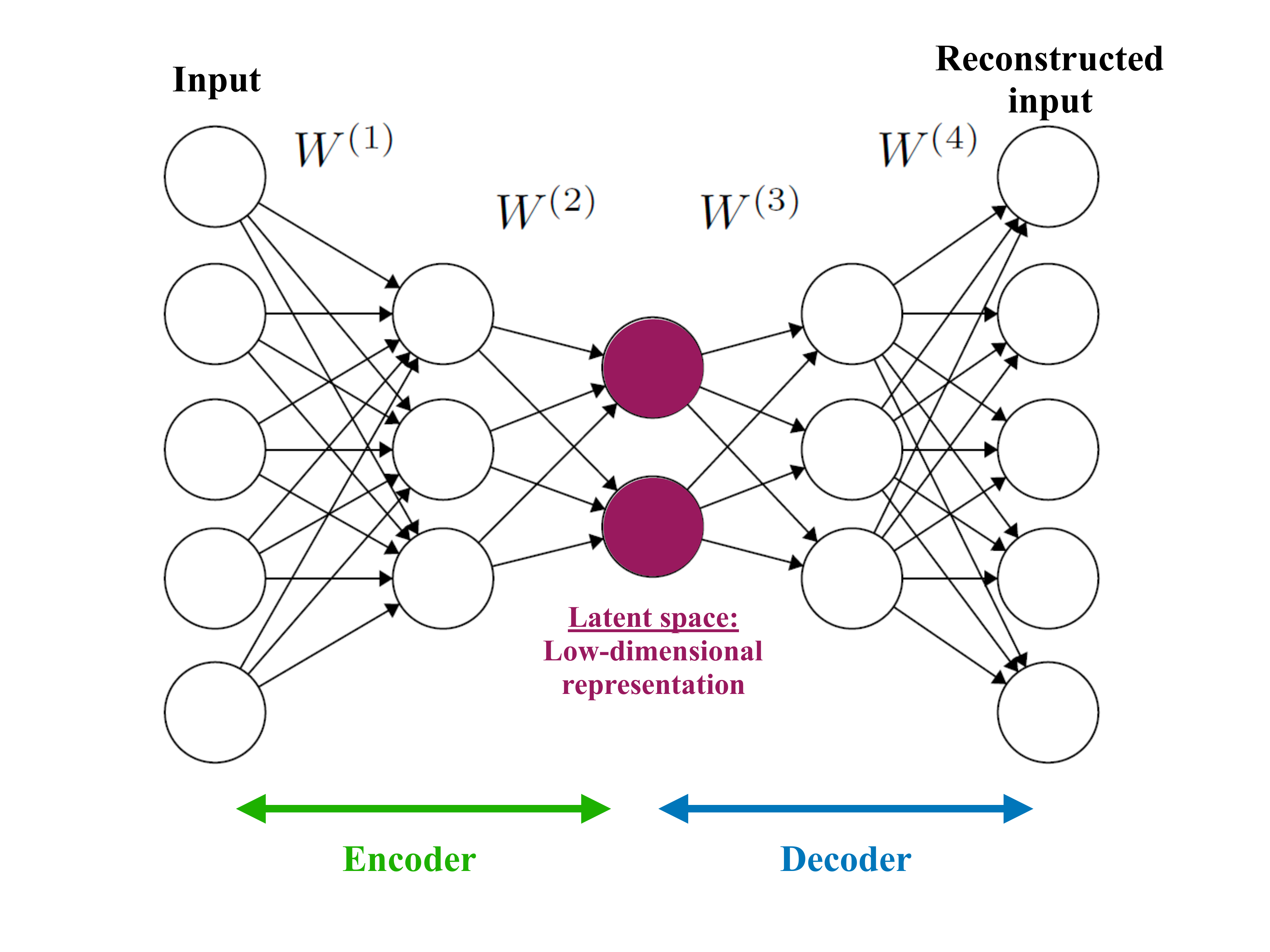}
\caption{A simplified example of an autoencoder architecture, used to perform compression, dimensionality reduction, and visualization. The network consists of two parts, the \emph{encoder} and the \emph{decoder}. The encoder reduces the dimensions of the input data, while the decoder reconstructs the input using the low-dimensional representation. The weights of the network are optimized during training to minimize the squared differences between the input and its reconstruction. The bottleneck of the network, also called the latent vector or latent space, represents the low-dimensional representation of the input dataset. }
\label{f:autoencoder_example}
\end{figure}

An autoencoder is a type of artificial neural network used to learn an efficient low-dimensional representation of the input dataset, and is used for compression, dimensionality reduction, and visualization \citep{gianniotis15, yang15, gianniotis16, maxu18, schawinski18}. Figure \ref{f:autoencoder_example} shows a simplified example of an autoencoder architecture. The network consists of two parts, the \emph{encoder} and the \emph{decoder}. In the encoding stage, the input data propagates from the input layers to the hidden layers, which typically have a decreasing number of neurons, until reaching the bottleneck, which is the hidden layer consisting of two neurons. In other words, the encoder performs a compression of the input dataset, by representing it using two dimensions. Then, the decoder reconstructs the input data, using the information from the two-dimensional hidden layer. The weights of the network are optimized during the training, where the loss function is defined as the squared difference between the input data and the reconstructed data. Once trained, the dataset can be represented in a low-dimensional space, also called the latent space, using the values given in the bottleneck hidden layer. As for all neural network-based algorithms, these networks are general and flexible, and can be used to represent very complex datasets. However, this complexity comes with a price. Such networks can have many different architectures, and thus a large number of free parameters, and are difficult to interpret (see lectures by M. Huertas-Company for additional details). 

\subsubsection{Self Organizing Maps}\label{sec:som}

\begin{figure}
\includegraphics[width=0.5\textwidth]{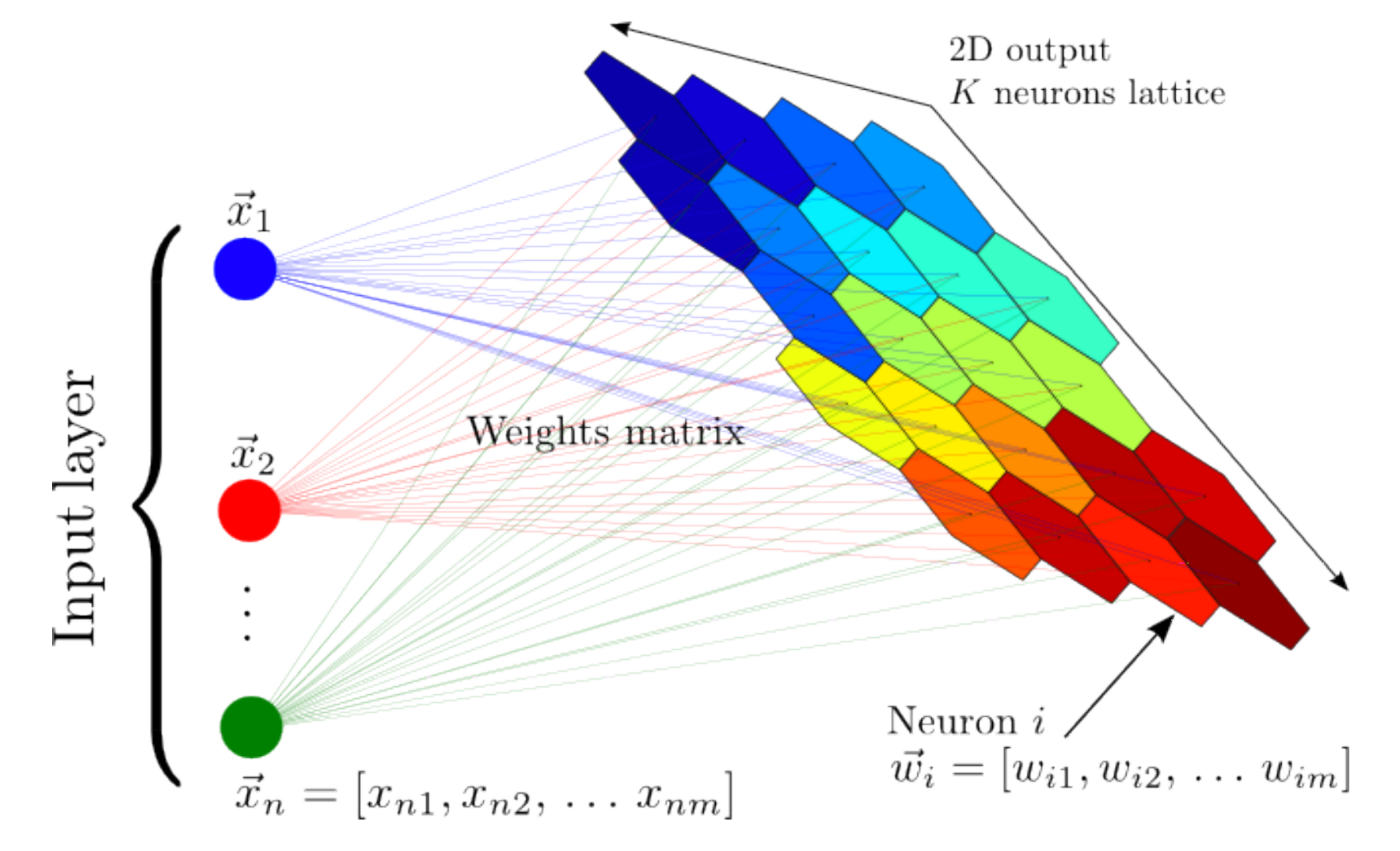}
\caption{A schematic illustration of a self-organizing map, taken from \citet{carrasco_kind14}. The input dataset consists of $n$ objects, each with $m$ features, and it is mapped to a two-dimensional lattice of $k$ neurons. Each neuron is represented by a weight vector, which has the same dimensions as the input objects. The weights are characteristics of the neurons themselves, and they represent the coordinate of each neuron in the input data space. These weights are updated during the training process, and once the algorithm is trained, they represent a set of templates (or prototypes) that describe the different objects in the dataset. }
\label{f:som_example}
\end{figure}

A self-organizing map (also named Kohonen map; \citealt{kohonen82}) is a type of artificial neural network that is trained in an unsupervised manner and produces a low-dimensional (typically two-dimensional) representation of the input dataset. During training, the two-dimensional map self-organizes itself to match the input dataset, preserving its topology very closely. In Astronomy, self-organizing maps have been used to perform semi-supervised classification and regression, clustering, visualization of complex datasets, and outlier detection (see e.g., \citealt{meusinger12, fustes13, carrasco_kind14, armstrong16, polsterer16, armstrong17, meusinger17, rahmani18}). Figure \ref{f:som_example} shows a schematic illustration of a self-organizing map, taken from \citet{carrasco_kind14}. The input dataset consists of $n$ objects with $m$ features each. The network consists of an input layer with $m$ neurons, and an output layer with $k$ neurons, organized as a two-dimensional lattice. The neurons from the input layer are connected to the output layer with weight vectors, which have the same dimensions as the input objects ($m$ in this case). Contrary to typical artificial neural networks, where the weights are used to multiply the input object values, followed by an application of an activation function, the weights of self-organizing maps are characteristics of the output neurons themselves, and they represent the "coordinates" of each of the $k$ neurons in the input data space. That is, the weight vectors serve as templates (or prototypes) of the input dataset. 

The training of a self-organizing map is a competitive process, where each neuron in the output layer competes with the other neurons to best represent the input dataset. The first step of self-organizing maps is a random initialization of the weights, where typically, the initial weights are set to be equal to randomly-selected objects from the input dataset. Then, the algorithm iterates over the objects in the input dataset. In each iteration, the algorithm computes the distance between the particular object and all the neurons in the output layer, using the euclidean distance between the object's features and the weight vectors that are associated with the neurons. Then, the algorithm determines the closest neuron to the input object, and updates its weight vector to be somewhat closer to the input object. The algorithm also updates the weight vectors of the neighboring neurons, such that closer neurons are updated more than farther neurons. The update magnitude is determined by a kernel function, usually a Gaussian, which depends on a \emph{learning radius} parameter. The update magnitude of all the neurons depends on a \emph{learning rate} parameter, where both the \emph{learning radius} and the \emph{learning rate} decrease with time. The self-organizing map converges after a number of iterations, and in its final form it separates the input dataset into groups of similar objects, which are represented by nearby neurons in the output layer. The final weights of the network represent prototypes of the different groups of objects, and they are usually used to manually inspect the dataset. Self-organizing map is implemented in {\sc python} and is publicly available on {\sc github}\footnote{\url{https://github.com/sevamoo/SOMPY}}.

\begin{figure*}
	\centering
\includegraphics[width=0.98\textwidth]{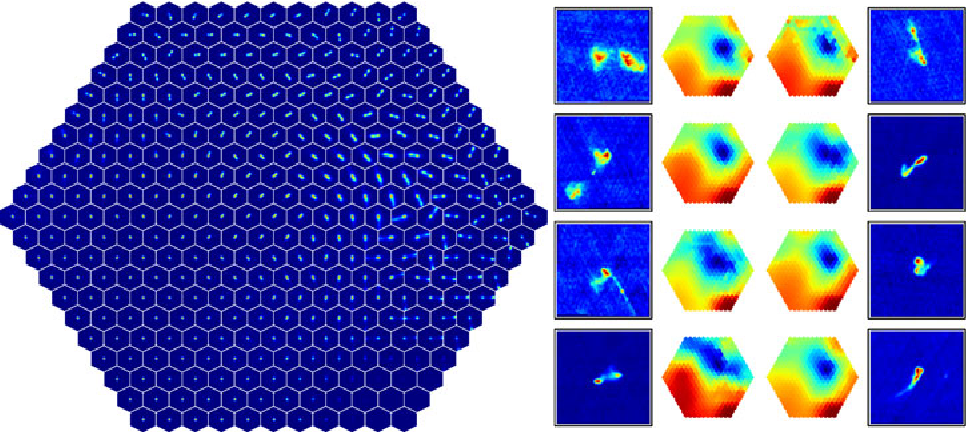}
\caption{Application of PINK to 200\,000 radio images from Radio Galaxy Zoo, taken from \citet{polsterer16}. The left panel shows the resulting two-dimensional map containing the derived prototypes. The right panel shows eight outliers that were selected based on their dissimilarity with the prototypes, and heatmaps that indicate their distance to all the prototypes. }
\label{f:pink_example}
\end{figure*}

Self-organizing maps are general and flexible, and their capability of sorting the input dataset onto a two-dimensional plane allows manual inspection of a relatively small number of prototypes, and use these to explore the structure of the dataset. However, in their simpler versions, self-organizing maps cannot be applied to astronomical images, since the algorithm is based on euclidean similarities, which is not invariant to rotations and flips. \citet{polsterer16} developed PINK, which is a self-organizing map that is invariant to rotations and flips, which is particularly useful for galaxy images, for example. Figure \ref{f:pink_example} shows an application of PINK to 200\,000 radio images, taken from \citet{polsterer16}. The left panel shows the resulting two-dimensional map, which contains the derived prototypes. These prototypes allow a clear separation of the input dataset into different morphological types, and by manually inspecting the map, one can explore the range of morphological types in the dataset, without manually inspecting thousands of images. The right panel shows eight outliers, which are objects which are not well-represented by any prototype.

\subsection{Anomaly Detection}\label{sec:outliers}

Outlier detection is the natural step after classification and the analysis of the dataset structure. In Astronomy, outlier detection algorithms were applied to various datasets, including galaxy and stellar spectra, galaxy images, astronomical light-curves such as variable stars, radio images, and more \citep{meusinger12, protopapas06, fustes13, nun16, agnello17, baron17, solarz17, hocking18, reis18a, segal18, giles19}. There are various types of outliers we expect to find in our datasets. Outliers can be objects that were not intended to be in our datasets, such as a quasar in a sample of stars, or various observational or pipeline errors. Such outliers are not scientifically interesting, but we wish to detect them and remove them from our samples. Outliers can be extreme objects drawn from the tail of well-characterized distributions, such as the most massive supermassive black hole or the most luminous supernova. Such objects are interesting because they allow us to test our models, which were built to describe the bulk of the population, in extreme regimes, and by studying them, we can learn more about the bulk of the population. The most interesting types of the outliers are the "unknown unknowns" \citep{baron17}, objects we did not know we should be looking for, and may be completely new objects which offer the opportunity to unveil new physics. Furthermore, in Astronomy, outliers can actually be very common phenomena, that occur on time-scales much shorter than other time scales of the system. For example, if every galaxy in the universe becomes green for 100 years, while in the rest of the time it evolves on a time scale of hundereds of million of years with its "regular" blue or red color, the probability of detecting a galaxy in its "green phase" using surveys such as the SDSS is close to zero. Therefore, although this "green phase" occurs in every galaxy and might be a fundamental phase in galaxy evolution, green galaxies will appear as outliers in our datasets. 

Unknown unknowns are usually detected serendipitously, when experts visually inspect their datasets, and discover an object that does not follow the accepted paradigm. Manual inspection becomes impractical in the big data era, where surveys provide millions of observations of a particular class of objects. Indeed, the vast majority of observations are no longer inspected by humans. To facilitate new discoveries, we must develop and use off-the-shelf algorithms to perform anomaly detection (see discussion by \citealt{norris17,norris17b}). Outlier detection is in some sense the ultimate unsupervised learning task, since we cannot define what we are looking for. Therefore, outlier detection algorithms must be as generic as possible, but at the same time they must be optimized to learn the characteristics of the dataset at hand, since outliers are defined as being different, in some sense, from the bulk of the population.

In most supervised and unsupervised tasks, the input dataset either consists of the original astronomical observations (spectra, light-curves, images, etc), or features that were extracted from the original observations, and there are advantages and disadvantages to both of these choices. However, anomaly detection should be carried out on a dataset which is as close as possible to the original dataset, and not on extracted features. First, defining features directly limits the type of outliers one might find. For example, in galaxy spectra, extracted features include the properties of the stellar population, and measurements of different emission lines. Such features will not carry information about new unidentified emission lines in a galaxy, and if such anomalous galaxies exist, they will not be marked as outliers. Second, extracted features are usually an output of some pipeline, which can sometimes fail and extract erroneous feature measurements. Such errors typically occur in outlying objects, since the models that are used to extract the features cannot describe them well. In such cases, the resulting features, which were wrongly measured, typically show values that are consistent with features measured for the bulk of the population, and such outliers cannot be detected (see \citealt{baron17} for examples). 

\subsubsection{Anomaly Detection with Supervised Learning}\label{sec:outliers_sup}

Supervised learning algorithms can be used to detect outliers, in both classification and regression tasks. When applied to new previously unseen objects, most supervised learning algorithms provide some measure of uncertainty or classification probability. Outliers can be defined as objects that have a large classification uncertainty or low classification probability. For example, a Random Forest algorithm is trained to distinguish between the spectra of stars and quasars. Then, it predicts the class (and class probabilities) of previously unseen objects. An object that is classified as a star with a probability of 0.55 (and probability of 0.45 to be a quasar) is probably more anomalous than an object that is classified as a star with a probability of 0.95. Therefore, the anomaly score of the different objects in the sample can be defined according to the classification probability. While easy to interpret, such a process will only reveal the outliers that "shout the loudest". Using again the star-quasar classification example, while a galaxy spectrum will be marked as an outlier by such a procedure, more subtle outliers, such as stars with anomalous metallicity, will not be detected. This is because supervised learning algorithms are optimized to perform classification (or regression), and as such, they will use only the features that are relevant for the classification task at hand. Therefore, objects which show outlier properties in these features will be detected, while objects that show outlier properties in features that are less relevant for the classification task, will not be detected. 

\subsubsection{Anomaly Detection with Unsupervised Learning}\label{sec:outliers_unsup}

There are several ways to perform outlier detection in an unsupervised setting. First, one can assign pair-wise distances between the objects in the sample, and define outliers as objects that have a large average distance from the rest of the objects (see e.g., \citealt{protopapas06, baron17}). As discussed in section \ref{sec:distances}, in some cases, using a euclidean metric might not result in an optimal performance, and one should consider other metrics. For example, the unsupervised Random Forest-based distance was shown to work particularly well on spectra \citep{baron17, reis18a, reis18b}, and cross correlation-based distances work well for time series \citep{protopapas06, nun16}. 

An additional way to perform outlier detection is by applying dimensionality reduction (which sometimes requires distance assignment as well). Once the high dimensional dataset is embedded onto a low dimensional plane, it can be visualized. Outliers can be defined as objects that are located far from most of the objects or on the edges of the observed distributions. The success of this process strongly depends on the procedure used to perform dimensionality reduction, and one must take into account the internal choices and loss function of the algorithm. For example, when using PCA to project the dataset onto a two-dimensional plane, one must take into account that while some outliers will be objects with extreme values in this 2D projection, and thus will be detected, other outliers can be objects that show extreme values in the other eigenvectors, which are not used for the projection and visualization, and thus will not show up as outliers. Another example is an auto-encoder, where some outliers will show up as extreme objects in the latent space (the two-dimensional representation), while other outliers will show typical values in the latent space, but a large reconstruction error on the decoder side. The final example is tSNE, where one must take into account the fact that the distances of the objects in the two-dimensional projection are not euclidean. In particular, while short distances in the original distance matrix are roughly conserved in the tSNE map, long distances are not. 

\subsubsection{One-Class SVM}\label{sec:ocsvm}

One of the most popular outlier detection algorithms is one-class SVM \citep{scholkopf99b}. In one-class SVM, the input data is considered to be composed of a single class, represented by a single label, and the algorithm estimates a distribution that encompasses most of the observations. This is done by estimating a probability distribution function which makes most of the observed data more likely than the rest, and a decision rule that separates these observations by the largest possible margin. This process is similar to the supervised version of SVM, but applied to a dataset with a single label. To optimize over the free parameters of SVM, such as the kernel shape and its parameters, the input dataset is usually divided into a training set and a validation set (there is no need for a test set, since this is an unsupervised setting). The algorithm is trained on the training set, resulting in some decision function, while the free parameters are optimized using the validation set. Therefore, the chosen kernel shape and its free parameters are chosen to give the highest classification accuracy on the validation set, where the classification accuracy is defined by the number of objects that are classified as inliers (the opposite of outliers) by the resulting decision function (see e.g., \citealt{solarz17}). The learned decision function is then used to define outliers. Outliers are objects that are outside the decision function, and their anomaly score can be defined by the distance of the outliers from the decision function. 

One-class SVM is feasible only with datasets composed of a handful of features. Therefore, it cannot be directly applied to astronomical observations such as images, light-curves, or spectra, but can be applied to photometry or derived features. One-class SVM is available in the {\sc scikit-learn} library\footnote{\url{https://scikit-learn.org/stable/modules/generated/sklearn.svm.OneClassSVM.html}}.

\subsubsection{Isolation Forest}\label{sec:iforest}

Another very popular outlier detection algorithm is Isolation Forest \citep{liu08}. Isolation Forest consists of a set of random trees. The process of building such a forest is similar to the training process of Random Forest, but here both the feature and the splitting value are \emph{randomly} selected at each node. Within each tree, outliers will tend to separate earlier from the rest of the sample. Therefore, the anomaly score can be defined as the depth at which a specific object was split from the rest, averaged over all the trees in the forest. The running time of Isolation Forest is $\mathrm{O}(N)$, where $N$ is the number of objects in the sample, and it can be applied to datasets with numerous features. \citet{baron17} compared its performance to that of the unsupervised Random Forest-based outlier detection, and found that Isolation Forest is capable of finding the most obvious outliers, those that "shout the loudest", but cannot detect subtle outliers, which are typically more interesting in an astronomical context. On the other hand, \citet{reis18a} and \citet{reis18b} found that when tuning the range of possible feature values that are randomly selected in each node (i.e., instead of defining the possible range to be between the minimum and maximum feature values, one could define the range to be between the 10th and 90th percentiles), Isolation Forest results in a comparable performance to that of the unsupervised Random Forest. Isolation Forest is available in the {\sc scikit-learn} library\footnote{\url{https://scikit-learn.org/stable/modules/generated/sklearn.ensemble.IsolationForest.html}}.

\section{Summary}

In recent years, machine learning algorithms have gained increasing popularity in Astronomy, and have been used for a wide variety of tasks. In this document I summarized some of the popular machine learning algorithms and their application to astronomical datasets. I reviewed basic topics in supervised learning, in particular selection and preprocessing of the input dataset, evaluation metrics of supervised algorithms, and a brief description of three popular algorithms: SVM, Decision Trees and Random Forest, and shallow Artificial Neural Networks. I mainly focused on unsupervised learning techniques, which can be roughly divided into clustering analysis, dimensionality reduction, and outlier detection. The most popular application of machine learning in Astronomy is its supervised setting, where a machine is trained to perform classification or regression according to previously-acquired scientific knowledge. While less popular in Astronomy, unsupervised learning algorithms can be used to mine our datasets for novel information, and potentially enable new discoveries. In section \ref{sec:open_questions} I list a number of open questions and issues related to the application of machine learning algorithms in Astronomy. Then, in section \ref{sec:further_reading}, I refer the reader to textbooks and online courses that give a more extensive overview of the subject. 

\subsection{Open Questions}\label{sec:open_questions}

The main issues of applying supervised learning algorithms to astronomical datasets include uncertainty treatment, knowledge transfer, and interpretability of the resulting models. As noted in section \ref{sec:unsup_forest}, most supervised learning algorithms are not constructed for astronomical datasets, and they implicitly assume that all measured features are of the same quality, and that the provided labels can be considered as ground truth. However, astronomical datasets are noisy and have gaps, and in many cases, the labels provided by human experts suffer from some level of ambiguity. As a result, supervised learning algorithms perform well when applied to high signal-to-noise ratio datasets, or to datasets with uniform noise properties. The performance of supervised learning algorithms strongly depends on the noise characteristics of the objects in the sample, and as such, an algorithm that was trained on a dataset with particular noise characteristics will fail to generalize to a similar dataset with different noise characteristics. It is therefore necessary to change existing tools and to develop new algorithms, which take into account uncertainties in the dataset during the model construction. Furthermore, such algorithms should provide prediction uncertainties, which are based on the intrinsic properties of the objects in the sample and on their measurement uncertainties.

The second challenge in applying supervised learning algorithms to astronomical datasets is related to knowledge transfer. That is, an algorithm that is trained on a particular survey, with a particular instrument, cadence, and object targeting selection, will usually fail to generalize to a different survey with different characteristics, even if the intrinsic properties of the objects observed by the two surveys are similar. As a result, machine learning algorithms are typically applied to concluded surveys, and rarely applied to ongoing surveys that have not yet collected enough labeled data. The topic of knowledge transfer is of particular importance when searching for rare phenomena, such as gravitational lenses in galaxy images, where supervised learning algorithms that are trained on \emph{simulated} data cannot generalize well to real datasets. This challenge can be addressed with \emph{transfer learning} techniques. While such techniques are discussed in the computer science literature, they are seldom applied in Astronomy.

The third challenge in applying supervised learning algorithms to astronomical datasets is related to the interpretation of the resulting models. While supervised learning algorithms offer an extremely flexible and general framework to construct complex decision functions, and can thus outperform traditional algorithms in classification and regression tasks, the resulting models are often difficult to interpret. That is, we do not always understand \emph{what} the model learned, and \emph{why} it makes the decisions that it makes. As scientists, we usually wish to understand the constructed model and the decision process, since this information can teach us something new about the underlying physics. This challenge is of particular importance in state-of-the-art deep learning techniques, which were shown to perform exceptionally-well in a variety of tasks. As we continue to develop new complex tools to perform classification and regression, it is important to devise methods to interpret their results as well.

When applying unsupervised learning algorithms to astronomical datasets, the main challenges include the interpretation of the results and comparison of different unsupervised learning algorithms. Unsupervised learning algorithms often optimize some internal cost function, which does not necessarily coincide with our scientific motivation, and since these algorithms are not trained according to some definition of "ground truth", their results might lead to erroneous interpretations of trends and patterns in our datasets. Many of the state-of-the-art algorithms are modular, thus allowing us to define a cost function that is more appropriate for the task at hand. It is therefore necessary to formulate cost functions that match our scientific goals better. To interpret the results of an unsupervised learning algorithm and to compare between different algorithms, we still use domain knowledge, and the process cannot be completely automatized. To improve the process of interpreting the results, we must improve the machine-human interface through which discoveries are made, e.g., by constructing visualization tools that incorporate post-processing routines which are typically carried out after applying unsupervised learning algorithms. Finally, as we continue to apply unsupervised learning algorithms to astronomical datasets, it is necessary to construct evaluation metrics that can be used to compare the outputs of different algorithms.

\subsection{Further Reading}\label{sec:further_reading}

To learn more about the basics of machine learning algorithms, I recommend the publicly-available machine learning course in {\sc coursera}\footnote{\url{https://www.coursera.org/learn/machine-learning}}. For an in-depth reading on statistics, data mining, and machine learning in Astronomy, I recommend the book by \citet{ivezic14}, which covers in greater depth many of the topics presented in this document, and many other related topics. For additional examples on machine learning in Astronomy, implemented in {\sc python}, I recommend astroML\footnote{\url{http://www.astroml.org/}} \citep{vanderplas12}.

\acknowledgments

I am grateful to I. Arcavi, N. Lubelchick, D. Poznanski, I. Reis, S. Shahaf, and A. Sternberg for valuable discussions regarding the topics presented in this document and for helpful comments on the text. 

\software{astroML \citep{vanderplas12},  
          scikit-learn \citep{pedregosa11}, 
          SciPy \citep{scipy01},
		  matplotlib \citep{hunter07},
		  and IPython \citep{perez07}.
          }

\bibliography{ref.bib}

\end{document}